\documentclass[aps,pre,groupedaddress]{revtex4}
\usepackage{graphicx}
\usepackage{amsmath}
\usepackage{bm}
\usepackage{epsf,epsfig,graphics}
\usepackage{float}
\usepackage{makeidx}
\usepackage{latexsym}
\usepackage{amssymb}
\usepackage{mathrsfs}
\usepackage{amsbsy}
\usepackage{subfigure}
\usepackage{CJKutf8}
\newcommand*{\vv}[1]{\vec{\mkern0mu#1}}

\usepackage{color}
\usepackage{morefloats}
\DeclareMathAlphabet\bfcal{OMS}{cmsy}{b}{n}
\usepackage{tabularx}
\usepackage{textcomp}
\usepackage{xcolor}
\usepackage{lineno}
\makeatletter
\usepackage{textcomp}
\begin{document}
\title{Steady-State Equilibrium and Nonequilibrium Noisy Network Dynamics}
\author{Pik-Yin Lai\footnote{pylai@phy.ncu.edu.tw}\begin{CJK*}{UTF8}{bsmi}(黎璧賢)\end{CJK*}}
\affiliation{Dept. of Physics and Center
for Complex Systems, National Central University, Chung-Li District, Taoyuan City 320317, Taiwan, R.O.C.
}
\affiliation{Physics Division, National Center for Theoretical Sciences, Taipei 106319, Taiwan, R.O.C.}
\date{\today}

\begin{abstract}
  The fluctuating dynamics of a network about its stable, noise-free steady state are theoretically investigated. Various causes of non-equilibrium dynamics are identified in terms of the properties and symmetry of the network connections and the noise covariance matrices.
  Several equivalent conditions are derived for the dynamics of the noisy network at equilibrium. In particular, non-equilibrium steady state (NESS) dynamics are analyzed in terms of the steady-state probability current and the drift velocity relative to the effective potential surface. Conventional physical Brownian  dynamics for overdamped fluctuating dynamics is analyzed from the perspective of the linearized fluctuating noisy network dynamics. Connection with the network reconstruction from time-series data is discussed. It is demonstrated that the overdamped Brownian dynamics in the physical system is a special case of the general noisy directed network in a NESS. Furthermore, a general fluctuation-dissipation relation is derived for the general non-equilibrium noisy network dynamics. These theoretical results are verified by numerical simulations.
\end{abstract}

\maketitle
\section{Introduction}
The study of complex systems as networks of interacting dynamical units has become a cornerstone of modern interdisciplinary science, with applications ranging from physical science, engineering, life science , neuroscience and ecology to social dynamics.
This has been boosted by the rapid increase in the vast volume of  data in a wide range of scientific, engineering, social and financial areas in the past three decades, which marked
the emergence of the Big-data era \cite{bigdata}.  The abundance and convenient availability of these data accessed through internet have enhanced the research activities of complex networks\cite{strogatz,newman,newmanbook}.
 A fundamental challenge in this field is understanding how such systems behave under the influence of internal or external noises and fluctuations. While the dynamics of isolated, noise-free systems are often tractable, the incorporation of stochasticity reveals a rich spectrum of behaviors, including non-equilibrium steady states (NESSs) characterized by persistent probability currents and broken time-reversal symmetry\cite{evans1993,Oono1998,Sekimoto2010,
 chiang2017electrical, MaPRE2017,chengLaiurns2021,
 LiaoPRE2022}. Nonequilibrium states are ubiquitous in biological and active matter systems, where energy input at the microscopic level drives  macroscopic fluxes. In contrast, for non-living systems composed of interacting entities modeled by network dynamics, often the internal structural, noise or dynamical symmetries are broken,which can lead to sustained currents in NESSs.

In many situations, the connections and interaction strengths in the a network are not known, but the dynamics of the nodes can be monitored. One of the challenging inverse problem is to retrieve the network connection weights or node-node interactions from the passively observed dynamics of the nodes that are accessible. In the network reconstruction problems developed in recent years\cite{
CLL2015,Hu2015}, with the help of noises, which accounts for external disturbances leading to fluctuating  dynamics in a network, it is possible to identify the network connectivity solely from the time-series dynamics of the nodes\cite{Lai2017,CT2017,TCL2018}. This noise-bridging approach, among many other methods, is part of a broader research area on network reconstruction problem\cite{ChengLai2022}.
 On the fundamental side, often the dynamical model of a complex network can be quite arbitrary or artificial, the  relationship between the associated fluctuating noisy dynamics and  possible physical non-equilibrium systems is still unclear and largely unexplored.
 
From the viewpoint of traditional statistical physics, the theoretical framework for analyzing stochastic dynamics has been heavily influenced by physical Brownian systems described by overdamped Langevin equations, which often assume equilibrium conditions obeying the Fluctuation-Dissipation relation (FDR) and detailed balance. In these classical scenarios, noise is typically tied to a universal temperature, and the steady-state distribution follows the equilibrium Boltzmann law. However, many real-world networks, such as neural circuits, gene regulatory networks, or synthetic coupled oscillators, operate far from equilibrium. Their connection topology can be directed, their noise sources can be heterogeneous and correlated, and their intrinsic dynamics are often nonlinear. These features break the conventional symmetry assumptions, leading to NESSs that cannot be described by equilibrium statistical mechanics.

This paper presents a unified theoretical framework for analyzing the fluctuating dynamics of general noisy networks about a stable steady state. We systematically bridge the gap between the well-established theory of physical Brownian motion and the broader class of stochastic dynamics on directed, nonlinear networks. A central theme is identifying the precise conditions under which a noisy network achieves equilibrium versus a NESS. We achieve this by examining the interplay between the network's connection matrix, the structure of the noise covariance matrix, and the resulting properties of the linearized dynamics matrix.
In this paper, we compare the general network dynamics under noise with the overdamped Langevin dynamics that models the Brownian-type physical systems. The scenarios of the usual Boltzmann equilibrium, equilibrium that differs from the Boltzmann distribution, and non-equilibrium steady-state (NESS) are discussed and distinguished. Furthermore, we identify various ingredients that give rise to the NESS in the noisy network dynamics.

The paper is organized as follows. In Section II, we introduce the general model of network dynamics under white noise and establish the foundational concepts of the steady-state distribution, effective potential, and probability flux. Section III is devoted to the analysis of linearized noisy network dynamics, where we derive the key relations for correlation matrices and discuss the connection with the framework of network reconstruction from time-series data. The conditions for equilibrium dynamics and the breaking of time-reversal symmetry are thoroughly examined in Section IV. In Section V, we place our results in a broader context by demonstrating how physical Brownian dynamics emerges as a special case of our network framework and derive a generalized FDR for NESS. The paper concludes with a summary and an outlook on future research directions. Several appendices provide technical derivations, including the solution of the Lyapunov equation and details on projected probability distributions.

\section{Network Dynamics Under white Noises}
 We consider general high-dimensional noisy dynamics governed by the stochastic equation of motion
 \begin{eqnarray}
{\dot {\vec x}}& =&{\vec {\cal F}}({\vec x})  +{\vec \eta}(t)
\label{xidot}
\end{eqnarray}
where ${\vec {\cal F}}({\vec x})$ is the force that drives the autonomous deterministic dynamics. 
${\vec \eta}(t)$ is  the zero-mean Gaussian white noise  given by:
\begin{equation}
\overline{{\vec \eta(t)} {\vec \eta^\intercal(t')}} = \boldsymbol{\sigma}\delta(t-t'); \qquad \overline{{\vec \eta(t)}}=0,\label{eta}
\end{equation}
where  $\boldsymbol{\sigma}$ is the symmetric covariance matrix and the $\overline{\cdots}$ stands for ensemble average over the noise.
A column vector is denoted by  $\vec{}$ , matrix quantities are denoted by bold fonts, and the transpose (of a vector or a matrix) is denoted by $^\intercal$.
 The stable fixed point(s) of noisy-free dynamics is denoted by ${\vec X}$, which is given by ${\vec {\cal F}}({\vec X})=0 $.
For the dynamics of the network considered here, we consider a network with $N$ nodes whose dynamics at the $i^{th}$ node is described by $x_i(t)$ and which is governed by the intrinsic node function $f_i(x_i;r_i)$ (in general non-linear), where $r_i$ denotes the parameter of the node. The force that acts on node $i$ is
\begin{equation}
{\cal F}_i({\vec x})=f_i(x_i;r_i)+\sum_{j\neq i}^N W_{ij} h(x_i, x_j),\label{Fi}
\end{equation}
where $W_{ij}$ is the connection weight from node $j$ to node $i$, and $h(x_i, x_j)$ is the coupling function for the interaction between nodes $i$ and $j$.
$h$ is a coupling function that describes the interaction between nodes $i$ and $j$. 

{\it Simulation of the Noisy Network node dynamics:-}
To examine the validity of the theoretical results, we carry out simulations by generating networks  with  a total of $N$ nodes and numerically solve stochastic differential equation \eqref{xidot} with noise given by \eqref{eta}.
 For the force acting on each node due to intrinsinc dynamics and network coupling governed by \eqref{Fi}),  we adopt the nonlinear logistic function $f_i(x_i)=r_ix_i(1-x_i)$unless otherwise stated, and  generate bi-directional  and directed weighted random (denoted by BWR and DWR respectively) Erdos-Renyi (ER) networks\cite{ER,random} with  edge connection probability $p$. The weights $W_{ij}$ of the edges are randomly chosen from a Gaussian distribution of mean $\mu_g$ and standard deviation $\sigma_g$.  The coupling function is taken to be $h(x_i,x_j)=x_j-x_i$ for simplicity. Each node $i$ of the entire network is subjected to an independent white noise with a diagonal noise matrix whose elements are given by $\sigma_{ij}=\sigma_{ii}\delta_{ij}$. In the simulations, the values $\mu_g=2$, $\sigma_g=2$ are used in the edge weights so that there are positive and negative weights.  In most cases the intrinsic dynamics is homogeneous for each node with $r_i=10$, and for the scenario of non-uniform $r_i$, its mean value  is also 10. In most cases, we simulate networks of sizes $N=100$  for typical  values of $p=0.2$ or $p=1$ (a fully connected network).  The network stochastic dynamics are solved by the Euler-Maruyama  method with a time step of $5\times 10^{-4}$.  After a sufficient time for achieving the steady state, the stochastic dynamics of ${\vec x}(t)$ for a duration $T$ (typically $10^4$) is generated for the measurement of various statistical averages and correlation functions.

\section{Steady-state distribution, effective potential and probability flux}
The general  noisy  dynamics described in (\ref{xidot}) and (\ref{eta}) can also be formulated in terms of the probability density $\psi({\vec x},t)$ and the probability flux  ${\vec J} $ by the Fokker-Planck equation:
\begin{eqnarray}
\partial_t \psi({\vec x},t)&=& -\nabla\cdot {\vec J},\quad {\vec J}=({\vec {\cal F}}-\frac{\boldsymbol{\sigma}}{2}\nabla)\psi,
 \end{eqnarray}
 where $\nabla$ is the $N$-dimensional gradient $\nabla_i\equiv \partial_{x_i}$.
 Assuming the dynamics will achieve a steady state (ss), with the steady-state distribution $\psi_{ss}$ which can be defined in terms of the effective potential function $\Phi$ as
 \begin{eqnarray}
    \psi_{ss}({\vec x})\sim e^{-\Phi({\vec x})}.\label{psissPhi}
\end{eqnarray}
Ensemble average of stochastic observable $B(\vec x)$ is denoted by 
$\langle B \rangle=\int B(\vec x) \psi_{ss}({\vec x}) d\vec x $,  which can be obtained in practice by
time-average over the asymptotic dynamics over an extended period of time.
In addition, denoting the (local) peak(s) of $ \psi_{ss}({\vec x})$ by $\vec x_{peak}$ and the steady-state mean of $\vec x$ by $\langle \vec x \rangle$. For general NESSs, $\vec X$, $\vec x_{peak}$ and $\langle \vec x \rangle$ can be nearby but all differ from each other due to the nonlinearity in $\vec{\cal F}$, and one can write  ${\vec x}_{peak}={\vec X} + {\vec \nu}$ and  $\langle \vec x \rangle={\vec X} + {\vec \mu}$.

 The steady-state  probability current $ {\vec J}_{ss}$ is given by
 \begin{eqnarray}
     {\vec J}_{ss}&=&\psi_{ss} {\vec V}_{ss},\quad \hbox{where } {\vec V}_{ss}={\vec {\cal F}}+\frac{\boldsymbol{\sigma}}{2}\nabla\Phi\label{Jss}
 \end{eqnarray}
 is the steady-state drift velocity. This suggests a force decomposition via
 \begin{equation}
{\vec {\cal F}}={\vec V}_{ss}-\frac{\boldsymbol{\sigma}}{2}\nabla\Phi.\label{Fdecomp}
 \end{equation}

 The divergence-free condition for the current in the steady state gives
 $\vec V_{ss}\cdot\nabla\Phi=\nabla\cdot \vec V_{ss} $ which leads to:
 \begin{eqnarray}
    \nabla\Phi\cdot(\vec {\cal F}+\tfrac{1}{2}\boldsymbol{\sigma}\nabla\Phi)&=& \nabla\cdot \vec {\cal F}+\tfrac{1}{2}\nabla\cdot (\boldsymbol{\sigma}\nabla\Phi)\\
  &=&  \hbox{Tr}( \nabla \vec {\cal F}+ \tfrac{1}{2} \nabla\nabla\Phi\boldsymbol{\sigma})
 .\label{varphi}
 \end{eqnarray}
 Eq. (\ref{varphi}) is a PDE for the effective potential gradient $ \nabla\Phi$ whose solution can in turn give $\psi_{ss}(\vec x)$.
 For NESS under a nonlinear force, the shift in the potential minimum from $\vec X$ will also lead to an extra component in the drift velocity that flows off the equi-$\Phi$ surface\cite{KwonAo2011}, giving
 \begin{equation}
{\vec V}_{ss}=-\frac{\boldsymbol{\alpha}}{2}\nabla\Phi+{\vec v}_{off}, \quad\boldsymbol{\alpha}=-\boldsymbol{\alpha}^\intercal.\label{Vssdecomp}
 \end{equation}
 The antisymmetric $\boldsymbol{\alpha}$ ensures that the first term in (\ref{Vssdecomp})  is normal to $\nabla\Phi $ and hence lies on the equi-$\Phi$ surface, and ${\vec v}_{off}$ denotes the drift that flows off the equi-$\Phi$ surface. Such a flow off from the equi-$\Phi$ surface has been reported in a nonlinear autonomous Brownian gyrator\cite{chang2021autonomous}.
 
\begin{figure}[H]
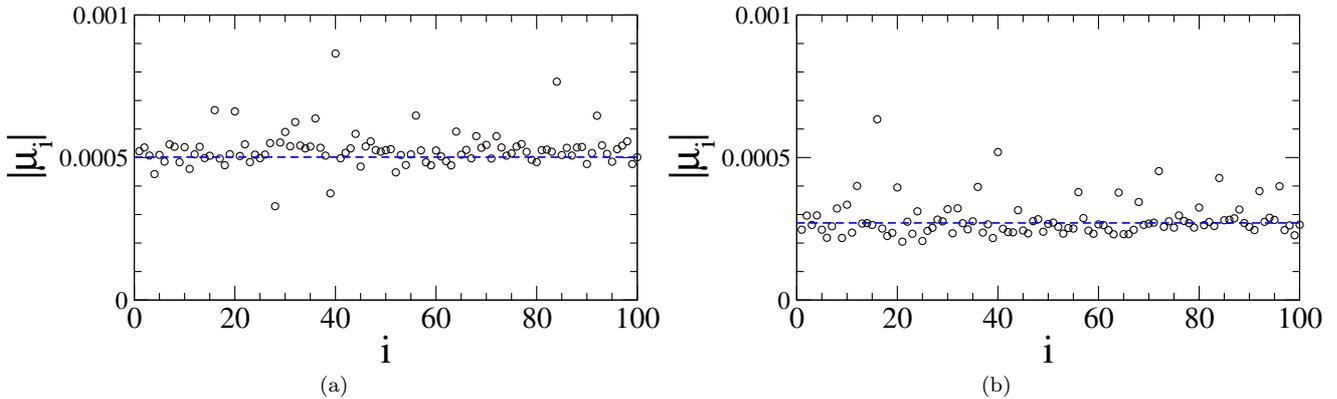

    \centering
     \subfigure[]{\includegraphics*[width=.48\columnwidth]{Fig1a.eps}}
     \subfigure[]{\includegraphics*[width=.48\columnwidth]{Fig1b.eps}}
\caption{Magnitude of the deviation of $\langle x_i  \rangle$ from $X_i$ for all the $N=100$ nodes. Network is under a diagonal and non-uniform noise matrix.(a) BWR network with $r=2$  (b) DWR network with $r=10$.  The measured $\langle \delta x_i^2  \rangle$ average over all $N$ nodes is marked by the horizontal dashed line. }
\label{mui}
\end{figure}

In practice, the deviation or fluctuation of $x_i$ about its mean value, $\delta x_i\equiv x_i- \langle x_i \rangle$ (hence  $\langle \delta x_i \rangle\equiv 0$), is usually recorded in simulations or experiments. 
As mentioned, ${\vec X}$ and $\langle  {\vec x}\rangle$  do not coincide due to the nonlinear contribution in $\vec{\cal F}$, with $\langle{\vec x}\rangle={\vec X} + {\vec \mu}$. Hence by definition $\Delta \vec x=\delta \vec x+{\vec \mu}$ and $\vec \mu= \langle \Delta \vec x\rangle$. Also from the analysis of the linearized dynamics in the next section(see \eqref{Deltaxt}), $ \langle \Delta \vec x\rangle=0 $ to linear order, thus $ \vec \mu\sim{\cal O}(\delta x ^2) $.
Fig. \ref{mui} plots the measured  $\mu_i$ from the simulations of bi-directional and directed networks of different strengths of the noise-free fixed point stability. The corresponding variances of the fluctuations, averaged over all nodes, are also measured (horizontal dashed line), verifying the claim that  $ |\vec \mu |\sim{\cal O}(\delta x ^2) $. To summarize,  $ \Delta{\vec x}$ and $\delta{\vec x}$ in general differ by ${\cal O}(\delta x ^2)  $ due to nonlinearity. In experiments or simulations, $\langle  {\vec x}\rangle$ (and hence  $\delta{\vec x}$) can be conveniently measured by the time-average of the time-series data of the steady-state dynamics, ${\vec x}(t) $.

Although $ \delta{\vec x}$ and $\Delta{\vec x}$  in general differ by $ {\cal O}(\delta x^2)$  for nonlinear steady-state dynamics, their corresponding  time-lag correlation functions  (in the long-time limit for the steady-state) agree much better, as shown below :
\begin{eqnarray}
    {\bf K}_\tau&\equiv& \langle \delta{\vec x}(\tau)\delta{\vec x}^\intercal(0)\rangle \\
 &=&  \langle{\vec x}(\tau){\vec x}(0)^\intercal\rangle -  \langle{\vec x}(\tau)\rangle  \langle{\vec x}^\intercal(0)\rangle \label{ktau2} \\
  &=&  {\bf C}_\tau - {\vec \mu}{\vec \mu}^\intercal,\label{KCmu}
\end{eqnarray}
where $ {\bf C}_\tau \equiv \langle \Delta\vec x(\tau)\Delta\vec x^\intercal (0) \rangle$.
Thus $ {\bf K}_\tau $ and $ {\bf C}_\tau $  differ by $ {\cal O}(\delta x^4)$  for nonlinear steady-state dynamics, and hence  $ {\bf C}_\tau $ can be well approximated by $ {\bf K}_\tau $ beyond the linear regime.
This provides the theoretical basis for the network reconstruction method using solely time-series data\cite{CLL2015,Lai2017,CT2017}, as outlined in the next section.

 In general, the interplay of nonlinearity in  ${\vec{\cal F}}({\vec x})$ and the non-equilibrium condition gives rise to a shift of the local maximum of $\psi_{ss}({\vec x})$, ${\vec x}_{peak}$, from the nearby noise-free fixed point ${\vec X}$\cite{KwonAo2011},  ${\vec x}_{peak}={\vec X} + {\vec \nu}$. This is demonstrated by the simulation results in Appendix F.
 Only in two special cases, to be discussed in the following, does the local probability maximum coincide with the noise-free fixed point, i.e. $\vec \nu=0 $.

\subsection{Equilibrium case}
The first special case is the equilibrium case, in which the steady-state current vanishes. From (\ref{Fdecomp}), one has 
\begin{equation}
\nabla\Phi=-2\boldsymbol{\sigma}^{-1}\vec{\cal F}.\label{gradPhieqm}
\end{equation}
Hence, the equilibrium fixed point ${\vec X}$ satisfying ${\cal F}({\vec X})=0$ is also a saddle point of $\Phi$, and a stable ${\vec X}$ is also the peak location of $\Phi$, i.e. $\vec x_{peak}={\vec X}$. The equilibrium potential is given by
\begin{equation}
    \Phi(\vec x)=\Phi(0) -2\int_{{\vec 0}\to{\vec x} }\boldsymbol{\sigma}^{-1}\vec{\cal F}({\vec x'})\cdot d{\vec x'}\label{eqmPhi}
\end{equation}
provided that the line integral above is path independent, i.e., satisfying the equilibrium condition. For the  case  of a conservative force with ${\vec{\cal F}}({\vec x})=-\nabla U({\vec x})$, the equilibrium condition is satisfied only if  $\boldsymbol{\sigma}=\propto{\bf I}$ (say $\boldsymbol{\sigma}=2 D{\bf I}$),
 and the equilibrium probability distribution is Boltzmann: $ \psi_{eqm}(\vec x)\sim e^{-\frac{U(\vec x)}{D}}$. Notice that in general $\vec {\cal F}$ can be nonlinear and hence ${\vec X}=\vec x_{peak}\neq \langle\vec x\rangle$. The deviation of $ \langle\vec x\rangle$ from the noise-free equilibrium value, $\vec \mu=  \langle\vec x\rangle-\vec X$, approaches 0 in the low noise limit, and can be calculated using the large deviation theory.

\subsection{Linear $\vec{\cal F}(\vec x)$}
 The second is the special case of linear ${\vec{\cal F}}({\vec x}) $ (both $f_i(x_i)$ and the coupling function $h$ are linear in the network dynamics) with ${\vec{\cal F}}({\vec x}) ={\bf Q}\Delta\vec x$, where ${\bf Q}$ is a matrix independent of ${\vec x}$ and $\Delta{\vec x}\equiv{\vec x}-{\vec X}  $. Then from the steady-state solution (see (\ref{Deltaxt}) in next section), $\langle \Delta\vec x\rangle=0 $ and thus $\langle{\vec x}\rangle={\vec X}$. Also  since $\langle {\vec {\cal F}}({\vec x})\rangle = {\vec {\cal F}}(\langle{\vec x}\rangle)= {\vec {\cal F}}({\vec X})=0$, hence from \eqref{xidot}, $\langle\dot{\vec x}\rangle=0$.
With ${\vec{\cal F}}({\vec x}) ={\bf Q}\Delta\vec x$, (\ref{varphi}) reads
 \begin{equation}
      \nabla\Phi\cdot ({\bf Q}\Delta\vec x+\tfrac{1}{2}\boldsymbol{\sigma} \nabla\Phi)= \hbox{Tr}({\bf Q}^\intercal+ \tfrac{1}{2}\nabla  \nabla\Phi\boldsymbol{\sigma})\label{varphilin}
 \end{equation}
  and the effective potential can be solved using the Gaussian ansatz of $\Phi=\tfrac{1}{2}\Delta\vec x^\intercal {\bf C}^{-1}_0 \Delta\vec x$, where $ {\bf C}_0=\langle \Delta\vec x\Delta\vec x^\intercal\rangle$. This can be seen by direct substitution\cite{chengLaiurns2021} of the Gaussian ansatz in (\ref{varphilin}) which leads to Eq. (\ref{QCCQ}) ((\ref{QCCQ}) can also be derived directly from the solution (\ref{Deltaxt})) and further renders both sides of (\ref{varphilin}) to be zero. 
 Therefore, $\psi_{ss}(\vec x)$ also peaks at $\vec X=\langle{\vec x}\rangle=\vec x_{peak}$.
Furthermore, notice that the LHS of (\ref{varphilin}) is invariant if $\boldsymbol{\sigma}$ is replaced by $\boldsymbol{\sigma}+\boldsymbol{\alpha}$ for an antisymmetric matrix $\boldsymbol{\alpha}=-\boldsymbol{\alpha}^\intercal$, and the LHS of  (\ref{varphilin}) becomes
$ ({\bf C}^{-1}_0 \Delta\vec x )^\intercal ({\bf QC}_0+\frac{\boldsymbol{\sigma}+\boldsymbol{\alpha}}{2} ) ({\bf C}^{-1}_0 \Delta\vec x )=0$ for arbitrary $\Delta \vec x$. This implies ${\bf QC}_0 +\frac{\boldsymbol{\sigma}+\boldsymbol{\alpha}}{2}=0 $ and together with  (\ref{QCCQ}), one obtains (\ref{alpha}). One can also easily verify that the RHS of (\ref{varphilin}) also vanishes if $\boldsymbol{\sigma}$ is replaced by $\boldsymbol{\sigma}+\boldsymbol{\alpha}$.
 Hence for linear $\vec {\cal F}$, one has the general force decomposition $\vec {\cal F}= -\tfrac{1}{2}(\boldsymbol{\sigma} +\boldsymbol{\alpha)}\nabla\Phi$ in NESS. Such NESS results for the linear force case have also been derived in \cite{PNAS2005}.
 
\section{Linearized Noisy Network Dynamics}

Here we assumed that the system approaches some asymptotic dynamics in the presence of noises, and a stable noise-free fixed point $\vec X$ exists.  To describe the fluctuating dynamics  of the stochastic trajectory $x_i(t)$ about $\vec X$, one can expand (\ref{xidot}) up to linear order in $\Delta{\vec x}\equiv{\vec x}-{\vec X}  $ to give
 \begin{eqnarray}
\Delta{\dot{\vec x}} &=&{\bf Q}\Delta{\vec x} +{\vec \eta}(t)
,\quad{\hbox{where }} {\bf Q}\equiv \nabla \vec{\cal F}|_{\vec X}.\label{Deltaxdot}
\end{eqnarray}
One expects the range of validity of the linearization to be given by
 $|\Delta \vec x| \sim {\cal O}(\sqrt{\bar{\boldsymbol{\sigma}}})$, where
 $\bar{\boldsymbol{\sigma}}$ denotes the average of the non-zero elements of $\boldsymbol{\sigma}$.
For stable noise-free fixed points, the real parts of all eigenvalues of ${\bf Q}$ cannot be positive.
For steady-state noise-free dynamics, ${\bf Q}$ is time-independent, and  the linearized motion in (\ref{Deltaxdot}) can be solved to give
\begin{equation}
    \Delta{\vec x}(t)=e^{ t{\bf Q}}\Delta{\vec x}(0)+\int^t_0 dt' e^{ (t-t'){\bf Q}}{\vec \eta}(t') \xrightarrow{t\to\infty}\int^t_0 dt' e^{ (t-t'){\bf Q}}{\vec \eta}(t').\label{Deltaxt}
\end{equation}
Theoretical results relating the correlation function $ {\bf C}_\tau\equiv \langle\Delta{\vec x}(\tau) \Delta{\vec x}^\intercal(0)\rangle$ to ${\bf Q}$ and $\boldsymbol{\sigma}$ can be derived (see Appendix A for derivations) using  (\ref{Deltaxt}), 
\begin{eqnarray}
  {\bf C}_\tau&=&e^{\tau{\bf Q}} {\bf C}_0\label{Ctau}\\
  -\boldsymbol{\sigma}&=&{\bf Q}{\bf C}_0+{\bf C}_0 {\bf Q}^\intercal.\label{QCCQ}
\end{eqnarray}
Eq. (\ref{QCCQ}) is in the form of the Lyapunov equation for ${\bf C}_0$ and its (unique) solution is given by\cite{BehBH19}
\begin{equation}
{\bf C}_0=\int_0^\infty dt e^{t{\bf Q}} \boldsymbol{\sigma}e^{t{\bf Q}^\intercal}.\label{Lynsoln}
\end{equation}
 Furthermore, ${\bf C}_0$ can be solved analytically by spectral decomposing ${\bf Q}$, and the results are derived in Appendix B.
 
For the linearized dynamics, the effective potential in (\ref{psissPhi})
in the linearized regime reads\cite{PNAS2005}  $\Phi=\frac{1}2\Delta \vec x^\intercal{\bf C}_0^{-1}\Delta \vec x$.
$ \Phi(\vec x)$ can be calculated by perturbation theory beyond the linear regime as demonstrated in \cite{KwonAo2011}.

 Furthermore, the (linearized) force in (\ref{Deltaxdot}) be  decomposed as
\begin{eqnarray}
  \vec{\cal F}(\vec x)&=&{\bf Q}\Delta{\vec x}={\bf QC}_0\nabla \Phi\\
   &=&-\tfrac{1}{2}(\boldsymbol{\sigma}
   +{\bf\boldsymbol{\alpha}})\nabla \Phi
   \\
   \hbox{where }& &  -\boldsymbol{\alpha}={\bf Q} {\bf C}_0-{\bf C}_0 {\bf Q}^\intercal\label{alpha}
\end{eqnarray}
is anti-symmetric and can serve as a measure of the degree of non-equilibrium (see (\ref{KQQK}) for the equilibrium condition in the next section). 

It is worth-noting that for the linearized dynamics, \eqref{Deltaxdot} gives $\langle \Delta\vec x\rangle =0$, i.e. $\vec \mu=0$ , $\langle \vec x\rangle =\vec X$.  Hence $\Delta\vec x=\delta \vec x$ and $ {\bf K}_\tau= {\bf C}_\tau $. One can replace   $\Delta\vec x$ and ${\bf C}_\tau $ in Eqs. \eqref{Deltaxdot} to \eqref{alpha} respectively by $\delta \vec x$ and $ {\bf K}_\tau$ for the linearized dynamics near the neighbourhood of $\vec X$.

\subsection*{Network reconstruction from time-series data }
For network dynamics governed by (\ref{Fi}), the linearized fluctuating dynamics about the stable fixed point $\vec X$ is given  by (\ref{Deltaxdot}) with ${\bf Q}$  given by
\begin{equation}
     Q_{ij}= W_{ij}\partial_2h(X_i,X_j)+\left[  f_i'(X_i;r_i)+ \sum_mW_{im} \partial_1 h(X_i,X_m) \right]\delta_{ij}.\label{Qij}
\end{equation}
From the network reconstruction perspective, in practice $ {\bf K}_\tau $ is measured experimentally or in simulations using (\ref{ktau2}) by recording the time-series data of $x_i(t)$. Since  $ {\bf C}_\tau = {\bf K}_\tau $ in the linearized regime, (\ref{Ctau}) and (\ref{QCCQ}) give the reconstruction formulae for dynamics fluctuating around the linearized regime:
\begin{eqnarray}
  {\bf K}_\tau&=&e^{\tau{\bf Q}} {\bf K}_0
  ,\hbox{ and hence }  {\bf Q}=\frac{1}{\tau}\ln ( {\bf K}_\tau{\bf K}_0^{-1})\label{Ktau}\\
  \boldsymbol{\sigma}&=&-{\bf Q}{\bf K}_0-{\bf K}_0 {\bf Q}^\intercal.\label{FDR}
\end{eqnarray}
Eqs. \eqref{Ktau} and  \eqref{FDR} constitute the framework for network reconstruction for general  networks\cite{CLL2015,Hu2015,CT2017,Lai2017}.
(\ref{Ktau}) can be used to reconstruct ${\bf Q}$ and (\ref{FDR}) is  for the reconstruction of the noise matrix $\boldsymbol{\sigma}$.
Notice that (\ref{FDR}) implies that $({\bf Q}{\bf K}_0)_{ii}=-\frac{\sigma_{ii}}{2} $, and if $\boldsymbol{\sigma} $ is diagonal then $({\bf Q}{\bf K}_0)_{ij}=-({\bf Q}{\bf K}_0)_{ji}$ for $i\neq j$.

For the linear ${\cal F}({\vec x})$ case,  since  the whole phase space is the linearized regime, and one expects an even more accurate network reconstruction performance.
In addition, ${\bf Q} $ and $ \boldsymbol{\sigma} $ can also be efficiently reconstructed from the time series data by other methods, such as using stochastic force inference\cite{ChengLai2022}.

\section{Equilibrium Noisy Network Dynamics and Time-Reversal Symmetry}
Equilibrium is achieved when the fluctuations of the noisy network dynamics are time-reversal symmetric, i.e.,  $  {\bf K}_\tau= {\bf K}_{-\tau}$, which is equivalent to a symmetric time-lag correlation matrix $  {\bf K}_\tau= {\bf K}_\tau^\intercal $  for all $\tau$. In general, due to the nonlinearity of the force, ${\bf K}_\tau\neq {\bf C}_\tau$ even at equilibrium (see (\ref{KCmu})). However, one can show (see Appendix C) that the following are equivalent conditions for time-reversal symmetric dynamics (i.e., equilibrium)
\begin{eqnarray}
     {\bf K}_\tau&=& {\bf K}_\tau^\intercal  \Leftrightarrow  {\bf C}_\tau= {\bf C}_\tau^\intercal.
     \label{KKtrans}
   \end{eqnarray}  
   
\begin{figure}[H]
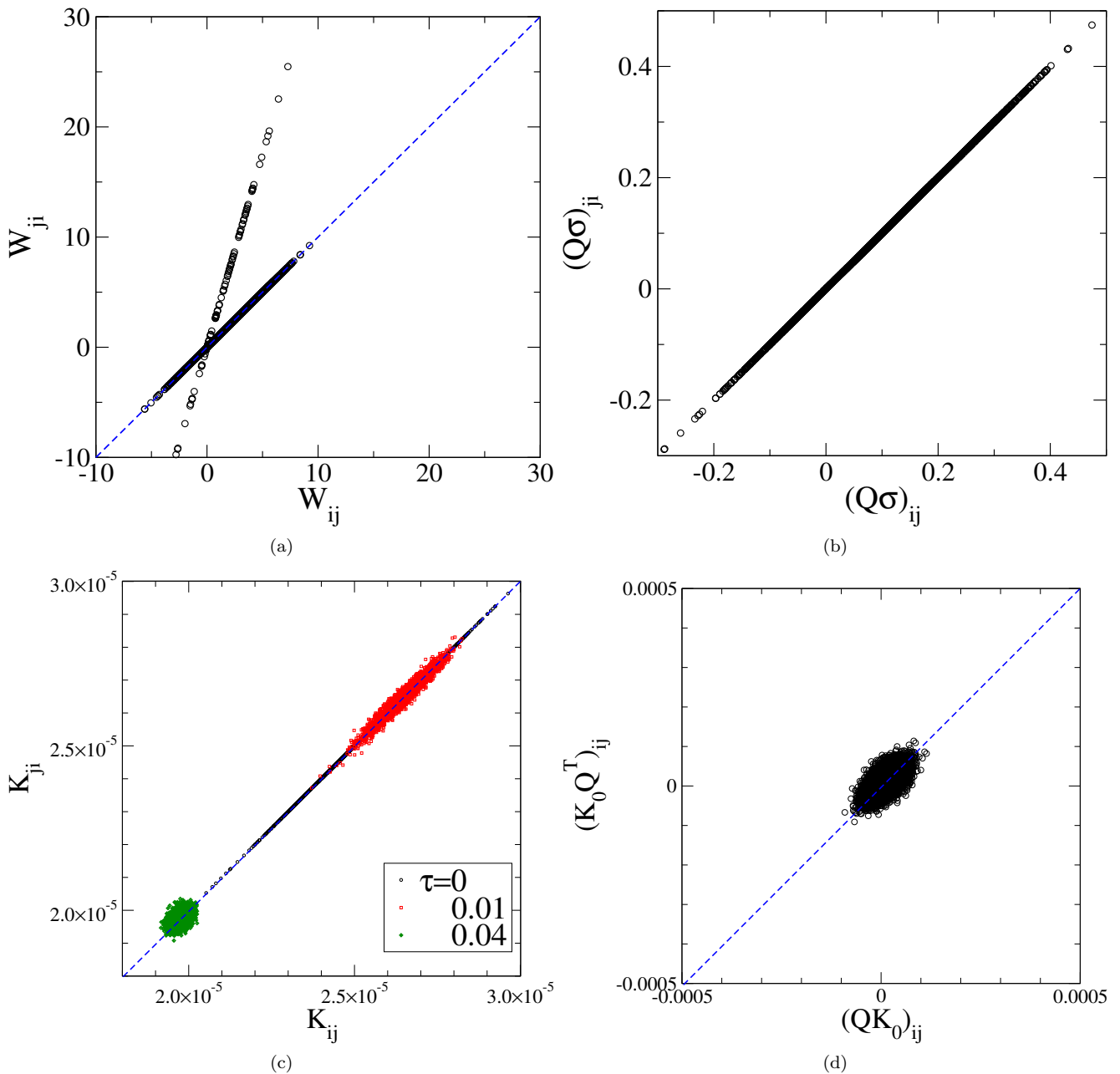

    \centering
        \subfigure[]{\includegraphics*[width=.48\columnwidth]{Fig2a.eps}}
        \subfigure[]{\includegraphics*[width=.48\columnwidth]{Fig2b.eps}}
        \subfigure[]{\includegraphics*[width=.48\columnwidth]{Fig2c.eps}}
        \subfigure[]{\includegraphics*[width=.48\columnwidth]{Fig2d.eps}}
       \caption{Linearized dynamics with asymmetric ${\bf Q}$ (i.e., non-conservative), but with non-trivial noise matrix adjusted so that (non-Boltzmann) equilibrium is still achieved. (a) $W_{ji}$ vs. $W_{ij}$ showing the asymmetric connection weights used. (b) Off-diagonal elements of ${\bf Q}\boldsymbol{\sigma}$ vs. its transpose
showing the noise is constructed to make ${\bf Q}\boldsymbol{\sigma}$ symmetric. (c) Off-diagonal elements of ${\bf K}_\tau$ vs. its transpose verifying the approximation in (\ref{KKtrans}).  (d)
Off-diagonal elements of ${\bf K}_0{\bf Q}^\intercal$ vs. ${\bf QK}_0$  verifying (\ref{KQQK}). 
}    \label{Qsigsymm} 
\end{figure}

   For the fluctuating dynamics in the linearized regime, in addition to \eqref{KKtrans}, the following equilibrium conditions are equivalent:
     \begin{eqnarray}
        {\bf Q} \boldsymbol{\sigma} &=&  \boldsymbol{\sigma}{\bf Q}^\intercal,\label{QsigsigQ}\\
     {\bf K}_0 {\bf Q}^\intercal&=& {\bf Q} {\bf K}_0= -\frac{\boldsymbol{\sigma}}{2}
    .\label{KQQK} \end{eqnarray}
(\ref{QsigsigQ}) implies time reversibility (and hence equilibrium dynamics) for certain noise and network structures. 
To demonstrate the equivalence of Eqs. (\ref{KKtrans})-(\ref{KQQK}), we prepare a network with asymmetry in ${\bf W}$ (Fig. \ref{Qsigsymm}a), but with a
diagonal $\boldsymbol{\sigma}$ adjusted such that the matrix ${\bf Q}\boldsymbol{\sigma}$ is symmetric (Fig. \ref{Qsigsymm}b). Noisy network dynamics simulations are performed, with nonlinear $\vec{\cal F}(\vec x)$ due to nonlinear $f_i(x_i)$, and the correlation functions $ {\bf K}_\tau $ are measured. (\ref{KKtrans}) is explicitly verified for various values of $\tau$ in Fig. \ref{Qsigsymm}c, and Fig. \ref{Qsigsymm}d shows that (\ref{KQQK}) holds. 
For the special case of symmetric ${\bf Q}\boldsymbol{\sigma}$ (equilibrium),  from \eqref{KQQK}, one gets
\begin{equation}
  \boldsymbol{\sigma}=-2   {\bf Q} {\bf K}_0.\label{FDRsym}
  \end{equation}

\begin{figure}[H]
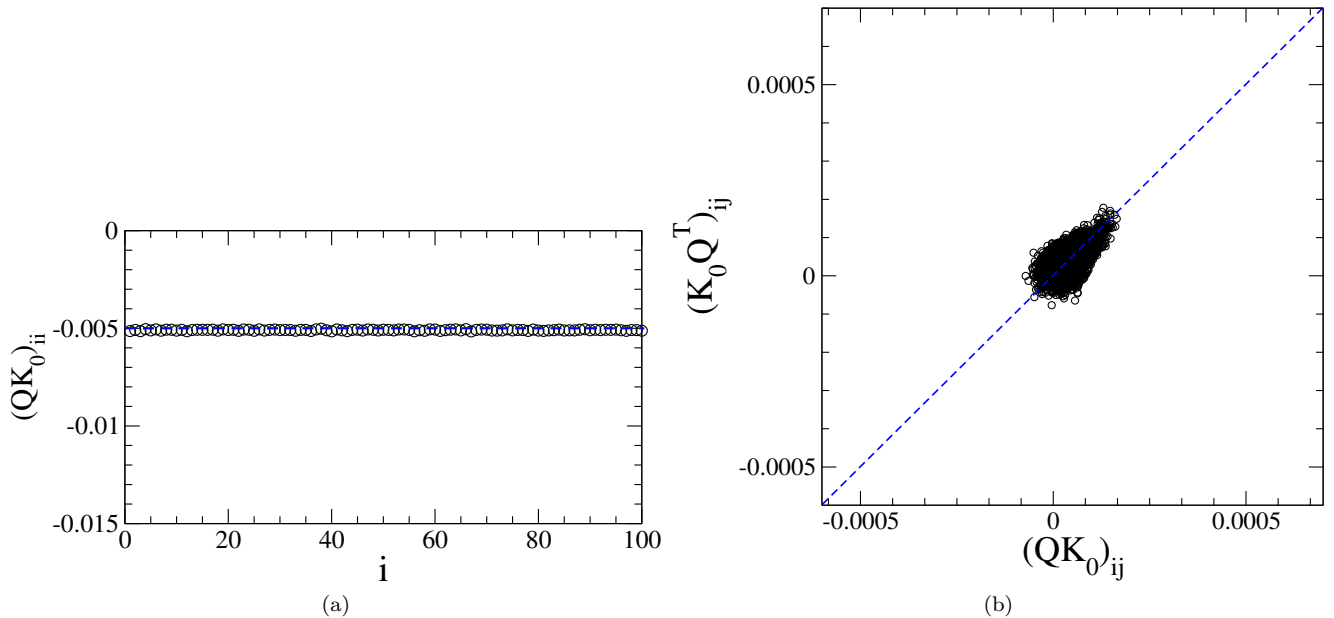

    \centering
     \subfigure[]{\includegraphics*[width=.48\columnwidth]{Fig3a.eps}}
\subfigure[]{\includegraphics*[width=.48\columnwidth]{Fig3b.eps}}
\caption{ (a) Diagonal elements of ${\bf QK}_0$. The horizontal line marks the value of $-\frac{\sigma}{2} $, confirming \eqref{QKsigI}. (b) Off-diagonal elements of ${\bf K}_0{\bf Q}^\intercal$ vs. ${\bf QK}_0$ for bi-directional weighted network with symmetric weights $W_{ij}$ and
noise $ \boldsymbol{\sigma}= \sigma {\bf I}$ verifying the noise reconstruction formula in (\ref{KQQK}) for undirected network, with $({\bf K}_0{\bf Q})_{ij}=({\bf QK}_0)_{ij}= 0$. }
\label{QK0}
\end{figure}

For bi-directional networks (symmetric $ {\bf Q}$) and uniform uncorrelated noises $\boldsymbol{\sigma}=\sigma{\bf I}$, 
one has 
\begin{equation}
 {\bf QK}_0= -\frac{\sigma}{2}{\bf I} ,\label{QKsigI}
\end{equation}
as verified in Fig. \ref{QK0}a. Fig. \ref{QK0}b confirms that the off-diagonal elements ${\bf K}_0{\bf Q}^\intercal={\bf QK}_0\approx 0$. Notice that in this case only ${\bf K}_0$ is needed to reconstruct the connection weight matrix $W_{ij}$\cite{CLL2015}.
However, for non-uniform $\boldsymbol{\sigma}$, ${\bf Q}\boldsymbol{\sigma} \neq \boldsymbol{\sigma}{\bf Q}=({\bf Q}\boldsymbol{\sigma}) ^\intercal$ even if
${\bf Q}={\bf Q}^\intercal$, and one needs to employ the method for reconstructing the directed network as given by (\ref{Ktau}).

\begin{figure}[H]
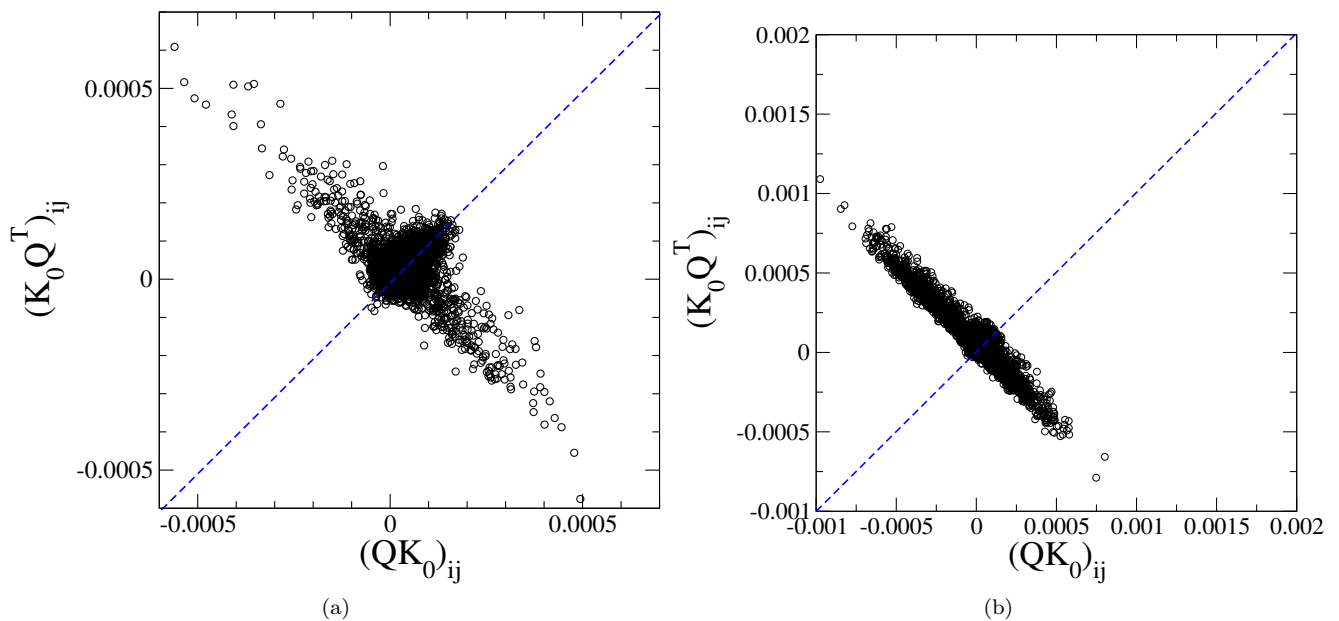

    \centering
 \subfigure[]{\includegraphics*[width=.48\columnwidth]{Fig4a.eps}}
 \subfigure[]{\includegraphics*[width=.48\columnwidth]{Fig4b.eps}}
    \caption{Off-diagonal elements of ${\bf K}_0{\bf Q}^\intercal$ vs. ${\bf QK}_0$ showing the time-reversal asymmetry in the NESS scenarios for (a) bi-directional weighted network
with symmetric weights $W_{ij}$ with non-uniform noise strengths $\sigma_{ij} = \sigma_{ii}\delta_{ij}$, . (b) Directed weighted network with asymmetric
weights and uniform noise strengths  $ \boldsymbol{\sigma}= \sigma {\bf I}$. Since    $ \boldsymbol{\sigma}$ is  diagonal in both cases, thus   $({\bf K}_0{\bf Q})_{ij}\simeq-({\bf QK}_0)_{ij}$.}
    \label{KQQKoffdiag} 
\end{figure}

In addition, notice that the RHSs of (\ref{FDR}) and (\ref{FDRsym}) (respectively for the NESS and equilibrium cases) are the same for diagonal elements. Hence, their differences will show up
only in the off-diagonal predictions of $\sigma_{ij}$. To distinguish them, one can plot the off-diagonal elements of ${\bf K}_0 {\bf Q}^\intercal $ vs.
$ {\bf Q} {\bf K}_0 $. The equilibrium case has been shown in Fig. \ref{QK0}b.   Fig. \ref{KQQKoffdiag} illustrates \eqref{KQQK} is indeed violated in the non-equilibrium scenarios.
 As shown in  Fig. \ref{KQQKoffdiag}, the off-diagonal elements of  ${\bf K}_0 {\bf Q}^\intercal $ and 
$ {\bf Q} {\bf K}_0 $ differ from each other significantly for the two NESS scenarios of a network
with symmetric weights $W_{ij}$ and  non-uniform diagonal noise strengths $\sigma_{ij} = \sigma_{ii}\delta_{ij}$ (Fig. \ref{KQQKoffdiag}a), and asymmetric
weights and uniform diagonal noise strengths  (Fig. \ref{KQQKoffdiag}b).

Under the equilibrium condition for noisy network dynamics, ${\bf Q}\boldsymbol{\sigma}$ is symmetric (which implies $\boldsymbol{\sigma}^{-1}{\bf Q}$ is also symmetric), $\psi_{eqm}({\vec x})$ obeys the Boltzmann distribution, then one can construct the potential as $ U({\vec x})=\frac{1}{2}{\Delta{\vec x}^\intercal\boldsymbol{\sigma}^{-1}{\bf Q}\Delta{\vec x}}$, which has a minimum at ${\vec X}$, i.e. ${\vec x}_{peak} =\vec X$.
On the other hand, the interplay of nonlinearity in  ${\vec{\cal F}}({\vec x})$ and the non-equilibrium condition gives rise to a shift of the local maximum of $\psi_{ss}({\vec x})$ from ${\vec X}$\cite{KwonAo2011}, and one has ${\vec x}_{peak}={\vec X} +  \vec \nu$ with $ \vec \nu\sim {\cal O}(\Delta x^2)$ as discussed in the previous section.

\section{Non-equilibrium Noisy Network Dynamics ($ {\bf Q \sigma} $ contains an asymmetric part)}
Therefore, from the results in the previous section, microscopic time-reversibility is broken for general noisy network dynamics even for the undirected network (${\bf Q} = {\bf Q}^\intercal$), unless $[{\bf Q},\boldsymbol{\sigma}] = 0$  (see Eq. (\ref{QsigsigQ})). Time-reversal symmetry is restored only for
the special case of a symmetric ${\bf Q}\boldsymbol{\sigma}$. Note that for this special case of ${\bf Q} \neq {\bf Q}^\intercal$ but ${\bf Q}\boldsymbol{\sigma}=\boldsymbol{\sigma}{\bf Q}^\intercal$, equilibrium state can be achieved with $\psi_{eqm} \propto e^{\Delta {\vec x}^\intercal\boldsymbol{\sigma}^{-1}{\bf Q}\Delta {\vec x} }$.

\begin{figure}[H]
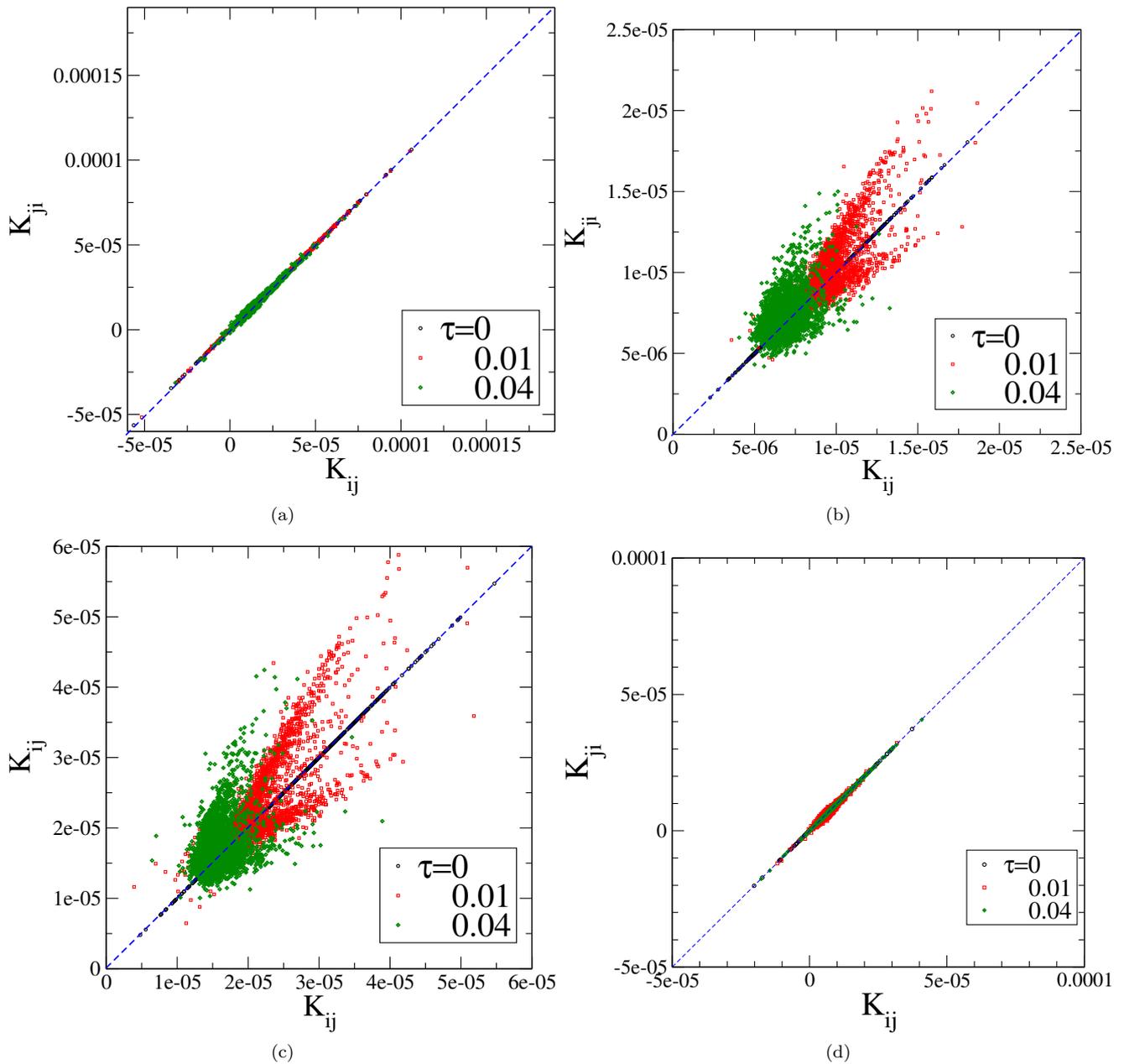

    \centering
\subfigure[]{\includegraphics*[width=.48\columnwidth]{Fig5a.eps}}
\subfigure[]{\includegraphics*[width=.48\columnwidth]{Fig5b.eps}}
\subfigure[]{\includegraphics*[width=.48\columnwidth]{Fig5c.eps}}   
     \subfigure[]{\includegraphics*[width=.48\columnwidth]{Fig5d.eps}}
\caption{Off-diagonal elements of ${\bf K}_\tau$ vs. its transpose for (a) bi-directional weighted network with symmetric weights $W_{ij}$ and
noise $ \boldsymbol{\sigma}= \sigma {\bf I}$  verifying the time-reversal symmetry for undirected network. (b) bi-directional weighted network with symmetric
weights $W_{ij}$ with non-uniform noise strengths $\sigma_{ij} = \sigma_{ii}\delta_{ij}$, and (c) directed weighted network with asymmetric weights and
uniform noise strengths $ \boldsymbol{\sigma}= \sigma {\bf I}$ . (d) Bi-directional weighted network with symmetric
weights $W_{ij}$ with uniform noise strength, $\boldsymbol{\sigma}\propto {\bf I}$ and the parameters $r_i$'s are non-uniform. Time-reversal symmetry holds only in (a) and (d). }
    \label{Kijoffdiag} 
\end{figure}

On the other hand, even for an undirected network, the noise has to be uniform
and (spatially) uncorrelated ($\boldsymbol{\sigma} \propto {\bf I}$) to preserve the time-reversal symmetry. Fig. \ref{Kijoffdiag} verifies explicitly that ${\bf K}_\tau$ is
symmetric in the case of an undirected network only under uniform diagonal noises (see Fig.  \ref{Kijoffdiag}a). If the noise strengths become
inhomogeneous, the symmetry is broken in general even if the network connection is symmetric (see Fig. \ref{Kijoffdiag}b). The physical picture
of the breaking of time-reversal symmetry under asymmetric ${\bf Q}$ or non-uniform noise strengths can be rationalized as
follows. Suppose node $j$ is under a strong noise and node $i$ is under a weak noise. The strong bombardment on $j$ will
result in an effective forcing on node $i$ via the coupling $Q_{ij}$, the effective “temperature” of $j$ is higher and thus
resulting in an unbalanced “heat flux” from $j$ to $i$ even if  $Q_{ij}=Q_{ji}$. Such an unbalanced driving or “heat flux” is the source
of time-reversibility breaking. Such an imbalance occurs trivially if ${\bf Q}$ is asymmetric (except if the noises are adjusted to make the matrix ${\bf Q}\boldsymbol{\sigma}$ symmetric). Detailed balance will be
broken in such a noisy network resulting in entropy production, even if the system achieves asymptotic dynamics of fluctuating about an attractor. The entropy production associated with each node can be theoretically calculated as the results will be presented in a separate paper. However, it should be noted that mere  non-uniform
intrinsic dynamics $f_i(x_i)$ (non-uniform external field/forcing) will not cause such a time-reversal symmetry breaking as shown in Fig. \ref{Kijoffdiag}d.
From  (\ref{QCCQ}) and (\ref{alpha}) , with ${\bf C}_0$ replaced by ${\bf K}_0$ in the linear regime, one gets
\begin{equation}
  {\bf Q}{\bf K}_0 =-\tfrac{1}{2}(\boldsymbol{\alpha}+ \boldsymbol{\sigma})
\end{equation}
revealing that the anti-symmetric $\boldsymbol{\alpha}$ caused the asymmetry in $ {\bf Q}{\bf K}_0 $ and thus gives rise to the
 non-equilibrium nature in the dynamics (see (\ref{KQQK})).

\subsection{Fluctuation-Dissipation Relation for NESS Network Dynamics}
\begin{figure}[H]
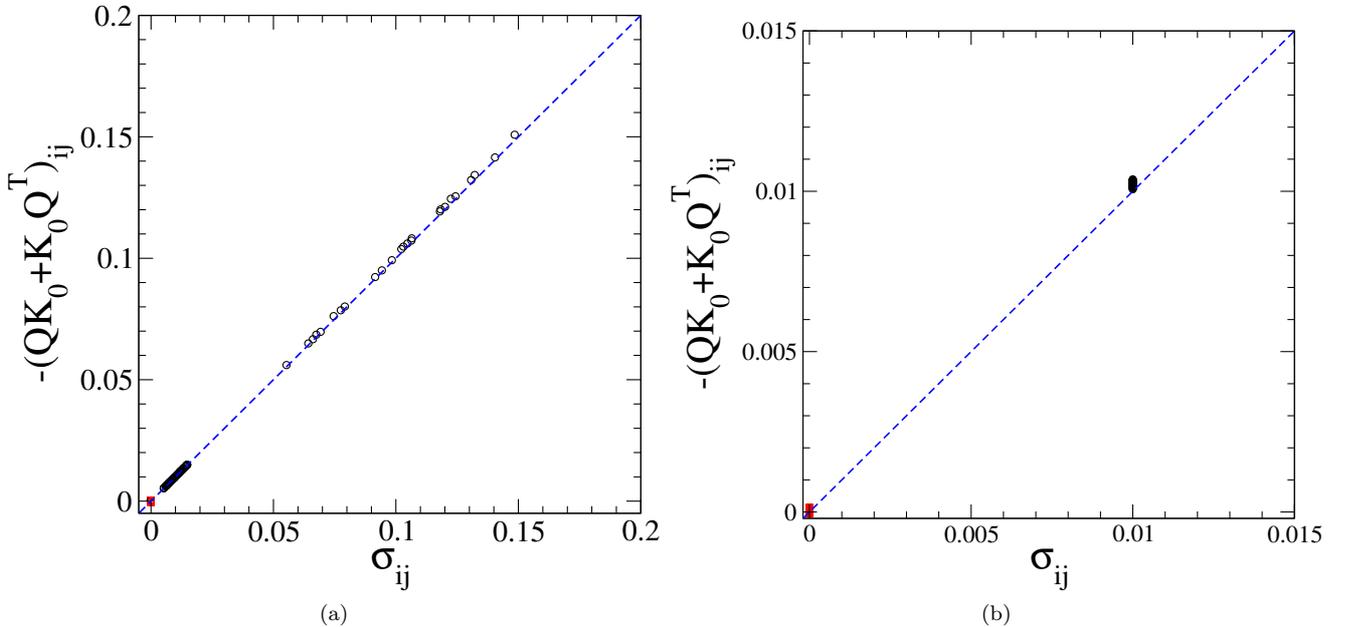

    \centering
    \subfigure[]{\includegraphics*[width=.48\columnwidth]{Fig6a.eps}}
     \subfigure[]{\includegraphics*[width=.48\columnwidth]{Fig6b.eps}}
    \caption{Elements of $-{\bf K}_0{\bf Q}-{\bf K}_0{\bf Q}^\intercal$ vs. $\boldsymbol{\sigma}$ to verify the FDR for network dynamics in the NESS scenarios for (a) bi-directional weighted network
with symmetric weights $W_{ij}$ with non-uniform noise strengths $\sigma_{ij} = \sigma_{ii}\delta_{ij}$,  (b) directed weighted network with asymmetric
weights and uniform noise strengths  $ \boldsymbol{\sigma}= \sigma {\bf I}$. }
    \label{QKKQsig} 
\end{figure}

Eq. \eqref{FDRsym} relates the noise strength and equilibrium correlation function, which is identified as the Fluctuation-Dissipation relation for equilibrium noisy network dynamics. Therefore, Eq. \eqref{FDR} can be viewed as a Fluctuation-Dissipation relation for NESS network dynamics. Fig. \ref{QKKQsig} plots the matrix elements of $-{\bf K}_0{\bf Q}-{\bf K}_0{\bf Q}^\intercal$ against the elements of $\boldsymbol{\sigma}$ using the measured ${\bf K}_0$ from simulations for two different scenarios of NESS dynamics: a network
with symmetric weights $W_{ij}$ and non-uniform noise strengths $\sigma_{ij} = \sigma_{ii}\delta_{ij}$(Fig. \ref{QKKQsig}a), and a network
with asymmetric weights and uniform diagonal noise strengths (Fig. \ref{QKKQsig}b). The Fluctuation-Dissipation relation \eqref{FDR}, for NESS dynamics,  is well confirmed.

\subsection{Distribution and dissipation in NESS network dynamics}
\begin{figure}[H]
    \centering
 \subfigure[]{\includegraphics*[width=.48\columnwidth]{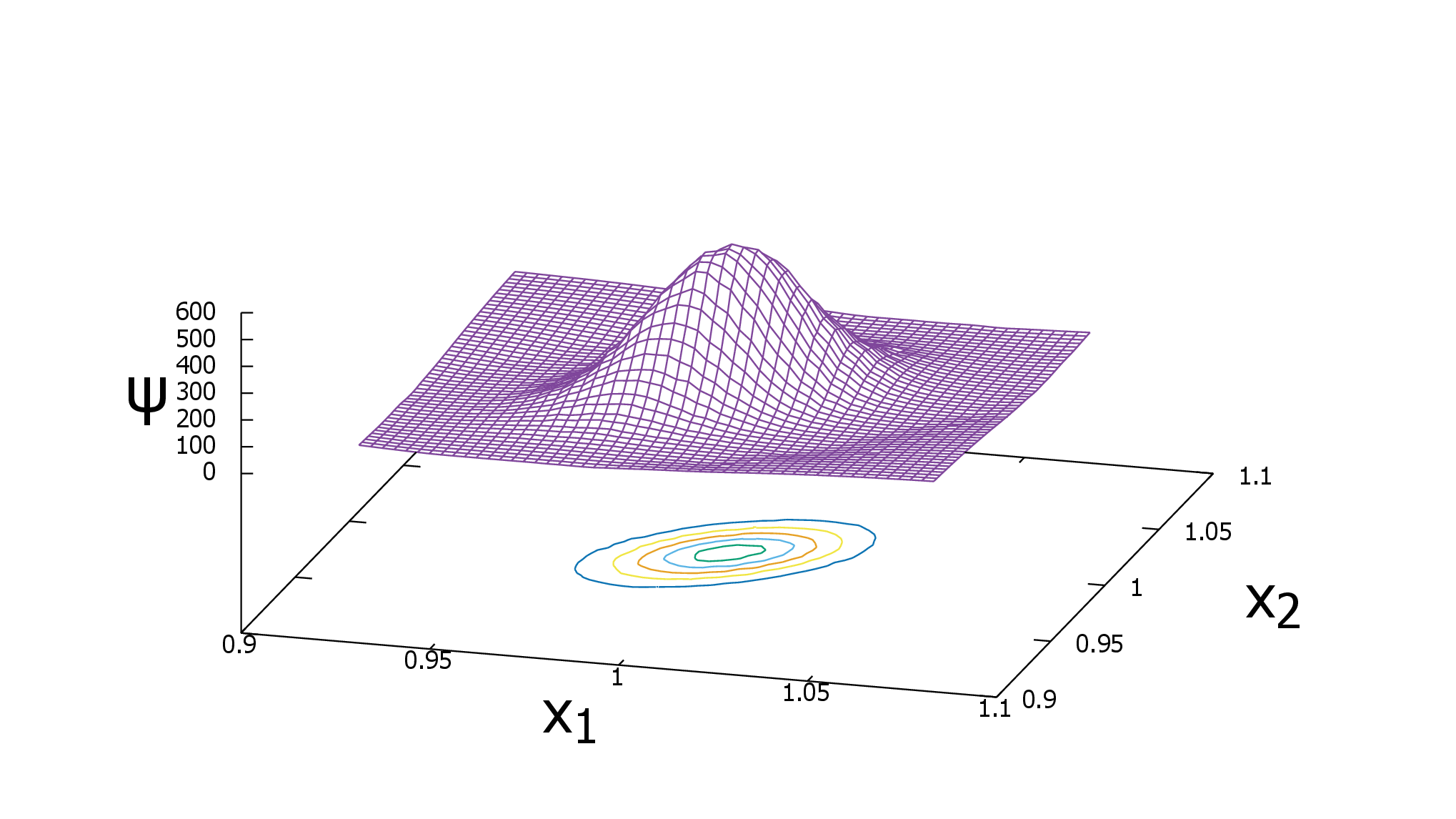}}
  \subfigure[]{\includegraphics*[width=.48\columnwidth]{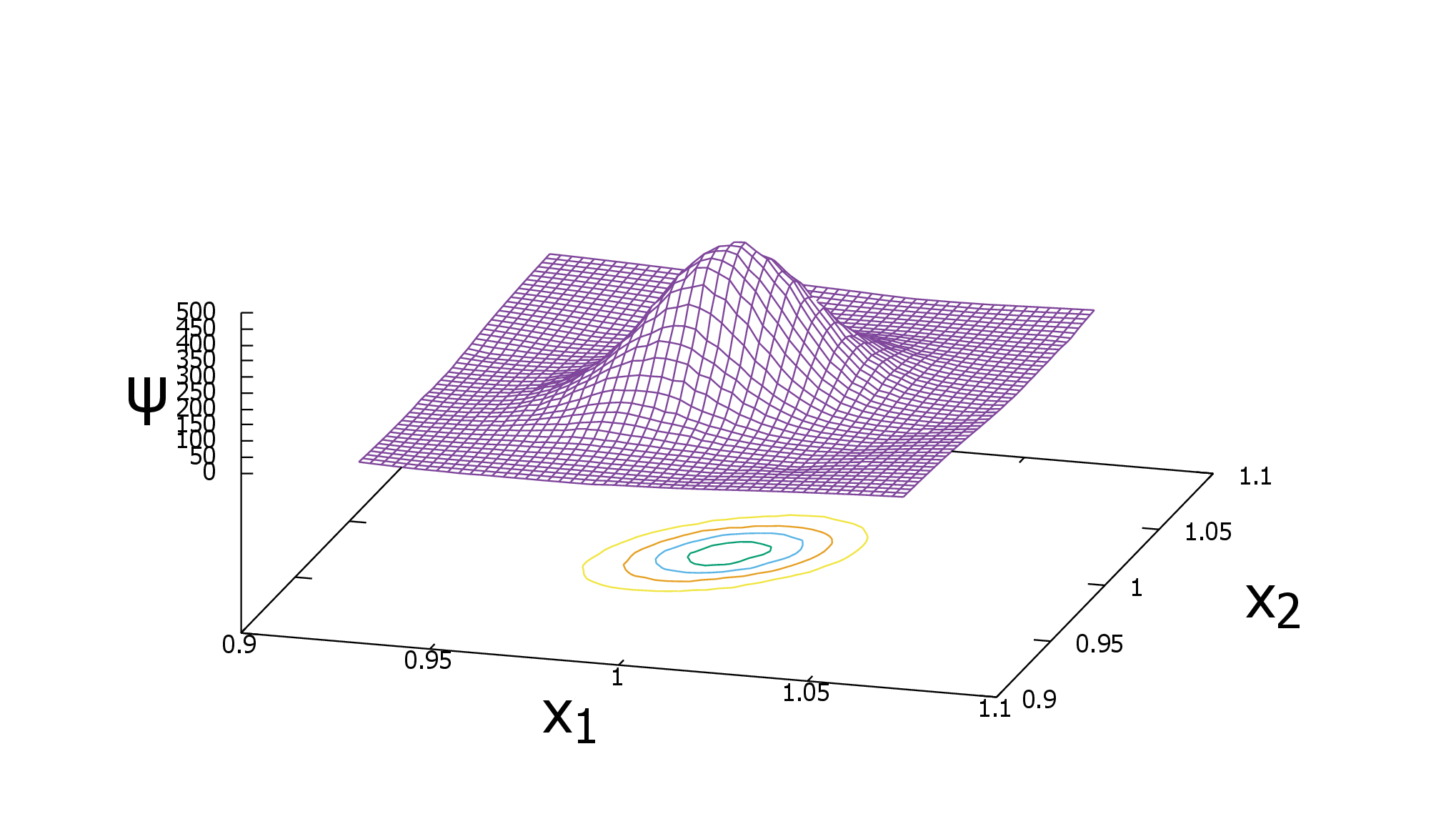}}
  \subfigure[]{\includegraphics*[width=.48\columnwidth]{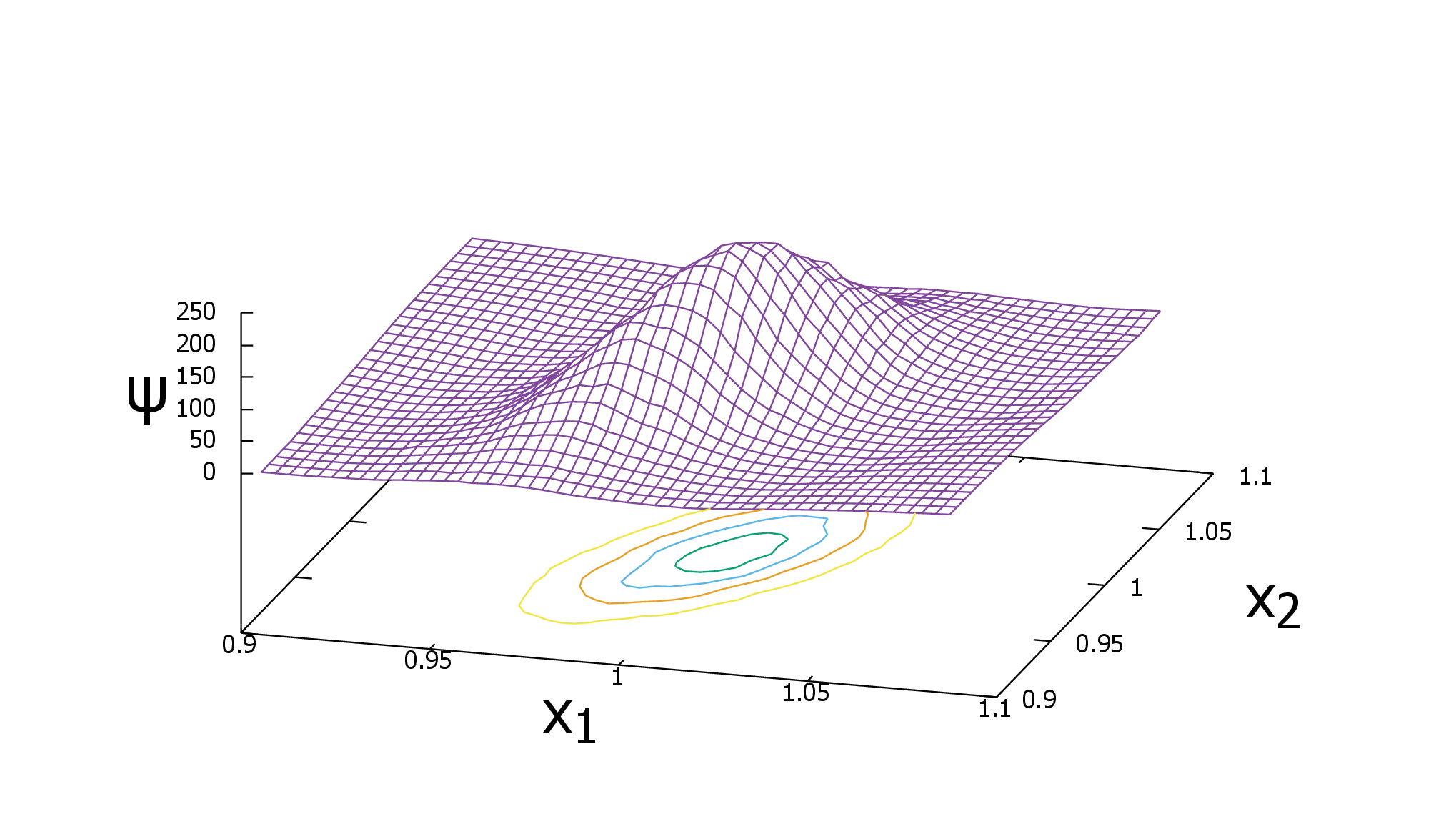}}
\subfigure[]{\includegraphics*[width=.48\columnwidth]{Fig7d.eps}}
    \caption{Dynamics from a two-node network to verify explicitly (\ref{psiss}) and (\ref{psieqm}) respectively for the NESS and equilibrium cases. With $\sigma_{11}= \sigma_{22} = 0.001$, and $r_1 = r_2 = 1$ unless otherwise stated. (a) $\psi_{\text{eqm}}(x_1,x_2)$ for the equilibrium case with $W_{12} = W_{21} = 1$. (b) $\psi_{\text{ss}}(x_1,x_2)$ for the NESS scenario with $W_{12} = 1$, $W_{21} = 0.2$. (c) $\psi_{\text{ss}}(x_1,x_2)$ for the NESS scenario with $W_{12} = 1$, $W_{21} = -1$. (d) The  measured values of $({\bf K}_0)_{11}$, $({\bf K}_0)_{22}$ and $({\bf K}_0)_{12}$, obtained from fitting the distributions with the Gaussian form, plotted against the theoretical values obtained from the solution  \eqref{FDR}.
    Five cases: $W_{12} = W_{21} = 1$; $W_{12} = 1$, $W_{21} = 0.2$; 
$W_{12} = -W_{21} = 1$;
   $W_{12} = W_{21} = 1$, 
   $\sigma_{22} = 0.01$; $W_{12} = 1$, $W_{21} = 0.2$,
    $\sigma_{22}=0.01$, $r_1=0.2$.
        }
    \label{2nodes} 
\end{figure}

 Under the assumption of fluctuation about a stable steady state (i.e., all real parts of the eigenvalues of ${\bf Q}$ cannot be positive), the system will approach a NESS\cite{PNAS2005}.  In general, fluctuations about the stable steady state, $ {\vec X}$, are always Gaussian but can be non-Boltzmann for directed networks and/or non-uniform $\sigma_{ij}$, giving rise to a NESS.
For general NESS, the linearized dynamics around $ {\vec X}$ can be described by the effective potential $\Phi=\frac{1}{2}\Delta{\vec x}^\intercal {\bf C}_0 ^{-1}\Delta{\vec x} $, and the steady-state probability distribution has the Gaussian form in the linearized regime
\begin{equation}
\psi_{\text{ss}}(\vec{x}) = \frac{e^{-\Delta \vec{x}^\intercal {\bf C}_0^{-1} \Delta \vec{x} / 2}}{(2\pi)^{N/2} \sqrt{\det {\bf C}_0}}.\label{psiss}
\end{equation}
In the linear regime, $\Delta{\vec x} $ and ${\bf C}_0 $ can be replaced by $\delta{\vec x} $ and ${\bf K}_0 $, which are measured in experiments or simulations in practice. Since it is not easy to show the high dimensional distribution in \eqref{psiss}, we show the simulation results 
 for a simple two-node network for illustrative purposes. Fig. \ref{2nodes} displays the simulation results of the steady-state distribution functions for equilibrium (Fig. \ref{2nodes}a) and NESS cases (Fig. \ref{2nodes}b and \ref{2nodes}c), verifying their Gaussian nature. The distributions are also fitting by the Gaussian form to obtain the elements of  ${{\bf K}_0}_{ij}$ and plotted against the
 theoretical values obtained from the solution of the Lyapunov equation \eqref{FDR} (Fig. \ref{2nodes}d), showing excellent agreement.
The present theoretical framework enables further understanding of stability and dissipation within the network's dynamics, as outlined below.
For general nonlinear $\vec{\cal F}$ and for dynamics beyond the linearized regime, it follows from the discussion in the previous section that the force acting on the nodes, in general, consists of three parts with 
\begin{eqnarray}{\vec{\cal F}}&=&-\frac{\boldsymbol{\sigma}}{2}{\bf C}_0 ^{-1}\Delta{\vec x}-\frac{\boldsymbol{\alpha}}{2}{\bf C}_0 ^{-1}\Delta{\vec x} +\hbox{higher order terms}.\label{decomp}\end{eqnarray}
The first part is the dissipative force pushing the system downhill in the potential landscape of $ \Phi({\vec x})=\frac{1}{2}\Delta \vec{x}^\intercal {\bf C}_0^{-1} \Delta \vec{x}+\hbox{higher order terms}$. The second part is a force that drives the motion/flow on
the equi-$\Phi$ surface, breaking the detailed balance. The third part consists of higher order terms due to the interplay of nonlinearity in $\vec {\cal F}(\vec x)$ and NESS giving rise to a probability current off the equi-$\Phi$ surface $\vec v_{off}$, as discussed in Section III.

For linearized  NESS, ${\bf C}_0$ and $\Delta \vec x$ can be replaced respectively by ${\bf K}_0$ and $\delta \vec x$, the distribution is given by (\ref{psiss}) with the steady-state probability current
 \begin{eqnarray}
     {\vec J}_{ss}&=&\psi_{ss} {\vec V}_{ss},\quad {\vec V}_{ss}=-\tfrac{1}{2}\boldsymbol{\alpha}{\bf K}_0^{-1}\delta{\vec x} ,\label{Vss}
 \end{eqnarray}
 and $\vec v_{off}$ is negligibly small, i.e., the presence of asymmetry in $ {\bf Q} {\bf K}_0 $ gives rise to a dominant contribution to the non-equilibrium flux.

\section{Brownian Dynamics from Noisy Network Dynamics Perspective}
Here we consider a physical system that can be described by Brownian dynamics in the overdamped regime. We first summarize the major results in Brownian dynamics, which  can be thought of as a physical system
of interacting Brownian colloids under a potential $U$ in a liquid medium. 

\subsection{Equilibrium case}
We first consider the overdamped Langevin equation describing the fluctuating Brownian dynamics  given by:
\begin{equation}
    -\boldsymbol{\zeta} \dot{\vec{x}} - \nabla U(\vec{x}) + \vec{F}_r(t) = 0,\label{Langevin}
\end{equation}
where $ \vec{F}_r(t)$ is the random force with the variance matrix governed by
\begin{equation}
    \langle \vec{F}_r(t) \vec{F}_r^\intercal(t') \rangle = 2k_B T \boldsymbol{\zeta} \delta(t-t').\label{FrFr}
\end{equation}
$x_i$’s are in general any physical quantities that depend on the phase space coordinates
and it is assumed that $x_i$ has a well-defined even or odd parity under time reversal.
The stochastic Brownian force specified by (\ref{FrFr}) already implicitly assumed the validity of the Fluctuation-Dissipation Theorem (FDT), namely the drag matrix is related to the covariance of the stochastic forces via $k_BT$, and  (\ref{Langevin}) and (\ref{FrFr}) describe the stochastic fluctuating dynamics in the overdamped regime. In addition,  the Onsager reciprocal relation ($\boldsymbol{\zeta} =\boldsymbol{\zeta} ^\intercal$) is also implied in (\ref{FrFr}), which is known to be valid in equilibrium and many non-equilibrium physical systems\cite{Doibook}.
The time-correlation function of fluctuations is time-reversal symmetric at equilibrium, in which FDT holds. FDT states that the drag on the particle and the random force the particle experiences
are related (two faces of the same coin), which follows from the same physical origin of stochastic bombardments on
the Brownian particle. The variance of the white noise given by $2k_BT\boldsymbol{\zeta} $ follows from the requirement that (\ref{Langevin}) and
(\ref{FrFr}) describe the equilibrium distribution of ~$\vec{x}$ that obeys the Boltzmann distribution.

From the network perspective, (\ref{Langevin}) can be written in the form of noisy network dynamics (\ref{xidot}) with $\vec {\cal F}=-\boldsymbol{\zeta}^{-1}\nabla U$ and $\boldsymbol{\sigma}=2k_BT\boldsymbol{\zeta}^{-1}$. One can then easily see that $\boldsymbol{\sigma}^{-1}\vec {\cal F}=-\nabla U/({2k_BT})$ and the Boltzmann equilibrium
with $\Phi(\vec x)=U(\vec x)/(k_BT)$ is verified (see (\ref{gradPhieqm}) ).

\begin{figure}[H]
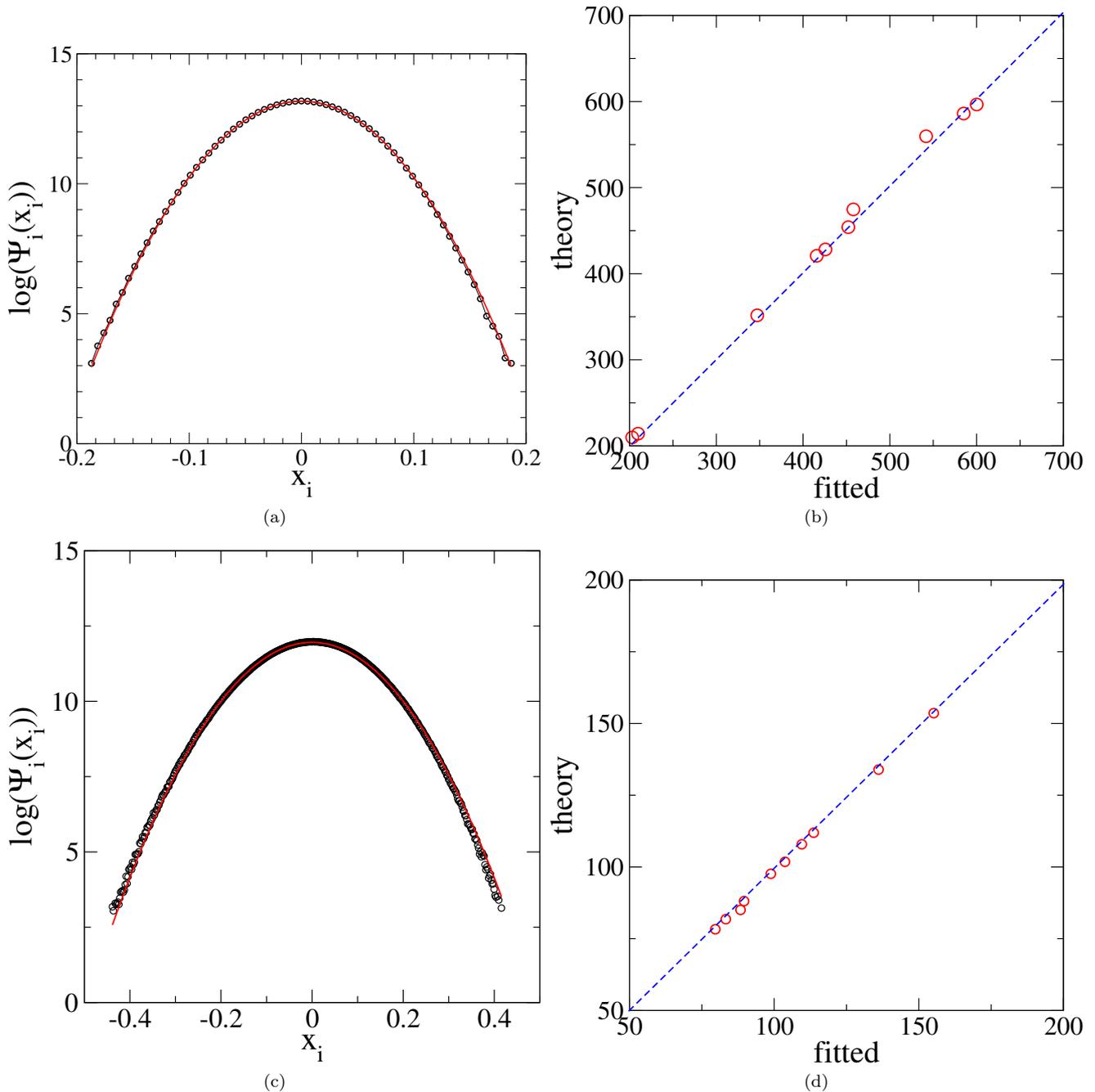

    \centering
\subfigure[]{\includegraphics*[width=.48\columnwidth]{Fig8a.eps}}
\subfigure[]{\includegraphics*[width=.48\columnwidth]{Fig8b.eps}}
\subfigure[]{\includegraphics*[width=.48\columnwidth]{Fig8c.eps}}
\subfigure[]{\includegraphics*[width=.48\columnwidth]{Fig8d.eps}}
    \caption{The reduced Gaussian distributions for (a) equilibrium and (c) NESS network dynamics. The theoretical width of the Gaussian distribution for  (b) equilibrium and (d) NESS network dynamics. The dashed straight line denotes the theoretical given by \eqref{Psiixi}.}    \label{logPsi2} \end{figure}

To investigate the fluctuation about the equilibrium, we consider the linearized dynamics about the equilibrium state (stable noise-free fixed point, $\vec{X}$,$\boldsymbol{\zeta}^{-1}\nabla U|_{\vec{X}}=0$), one obtains the network equation for $\Delta\vec{x}\equiv \vec{x}-\vec{X}$
\begin{equation}
\Delta \dot{{x}}_i = - \sum_k \sum_j {\zeta}^{-1}_{ij} {H}_{jk}
\Delta x_k + \sum_j {\zeta}^{-1}_{ij} {F}_{ri}(t),\label{Dxdot}
\end{equation}
where $H_{ij}\equiv \frac{\partial^2 U}{\partial x_i \partial x_j}|_{\vec{X}}$ 
is the Hessian matrix. The linear Langevin system with $U=\frac{1}{2}\Delta\vec{x}^\intercal{\bf H} \Delta\vec{x}$ (or equivalently linearized
the system described by (\ref{Dxdot}) ) fluctuates at equilibrium and follows Boltzmann distribution, thus the correlation matrix ${\bf C}_0$ or ${\bf K}_0$
 can be computed directly to give
\begin{equation}
{\bf H} =  k_B T  {\bf C}_0^{-1}= k_B T  {\bf K}_0^{-1},\label{HK0}
\end{equation}
which has been applied in dense colloidal systems\cite{Grunberg2004} to extract the effective spring network by measuring ${\bf K}_0\equiv \langle\delta\vec{x}\delta\vec{x}^\intercal \rangle$ from
direct observations. The Boltzmann distribution at equilibrium is  given by
\begin{eqnarray}
\psi_{\text{eqm}}(\vec{x})&\propto &e^{-\frac{U(\vec x)}{k_BT}} \\
& &={e^{-\frac{1}{2}\Delta \vec{x}^\intercal {\bf C}_0^{-1} \Delta \vec{x}}}= {e^{-\frac{1}{2}\delta \vec{x}^\intercal {\bf K}_0^{-1} \delta \vec{x}}}\quad \hbox{in the linearized regime}.\label{psieqm}
\end{eqnarray}
It is difficult to verify directly (\ref{psieqm}) since it is a very high-dimensional distribution, but one can measure the distribution
of the fluctuations of the $i^{th}$ node about its mean value $	\Psi_i( \delta {x_i})=\int\prod_{j\neq i} ^N d  \delta {x_i} \psi_{eqm}$. As shown in Appendix E, $\Psi_i( \delta \vec{x_i})$ is also Gaussian, which is verified in Fig. \ref{logPsi2}a. Also the width  of the Gaussian distribution depends on ${\bf K}_0$ and the node $i$ (see (\ref{Psiixi})), which is also verified in Fig. \ref{logPsi2}b for ten different
nodes in a network of 100 nodes.

From the viewpoint of linearized noisy network dynamics, (\ref{Dxdot}) is of the form
of network dynamics described in (\ref{Deltaxdot}), with
\begin{equation}
{\bf Q} = -\boldsymbol{\zeta}^{-1} {\bf H},\end{equation}
 From  relations (\ref{KQQK}) and (\ref{HK0}), we again verify that the  network variance matrix satisfies
 \begin{eqnarray}
\boldsymbol{\sigma}&=&
2k_B T\boldsymbol{\zeta}^{-1} \label{sigzeta}
\end{eqnarray}
 which agrees with the random force variance given by (\ref{FrFr}), giving the celebrated FDT at equilibrium.
Since (\ref{sigzeta}) follows from the network noise reconstruction formula (\ref{QCCQ}) or \eqref{FDR} under the equilibrium condition, thus (\ref{QCCQ}) or \eqref{FDR} can be viewed as a generalized Fluctuation-Dissipation Relation which holds for both equilibrium and non-equilibrium noisy network dynamics.
Note that 
even if both $\boldsymbol{\zeta}$ and ${\bf H}$ are symmetric, ${\bf Q}=-\boldsymbol{\zeta}^{-1}{\bf H}$ is still asymmetric in general unless $\boldsymbol{\zeta}$ and ${\bf H}$ commute.
Thus the Langevin Brownian system under a potential force is in general a directed network under white noise with variance given by (\ref{sigzeta}),
but still at equilibrium since ${\bf QK}_0 = {\bf K}_0{\bf Q}^\intercal$ (see (\ref{KQQK})). If $\boldsymbol{\zeta}\propto {\bf I}$ then ${\bf Q}\propto -{\bf H}$, resulting an undirected network.

\subsection{NESS and Fluctuation-Dissipation relation for
Brownian motion under a non-conservative force}
\begin{figure}[H]
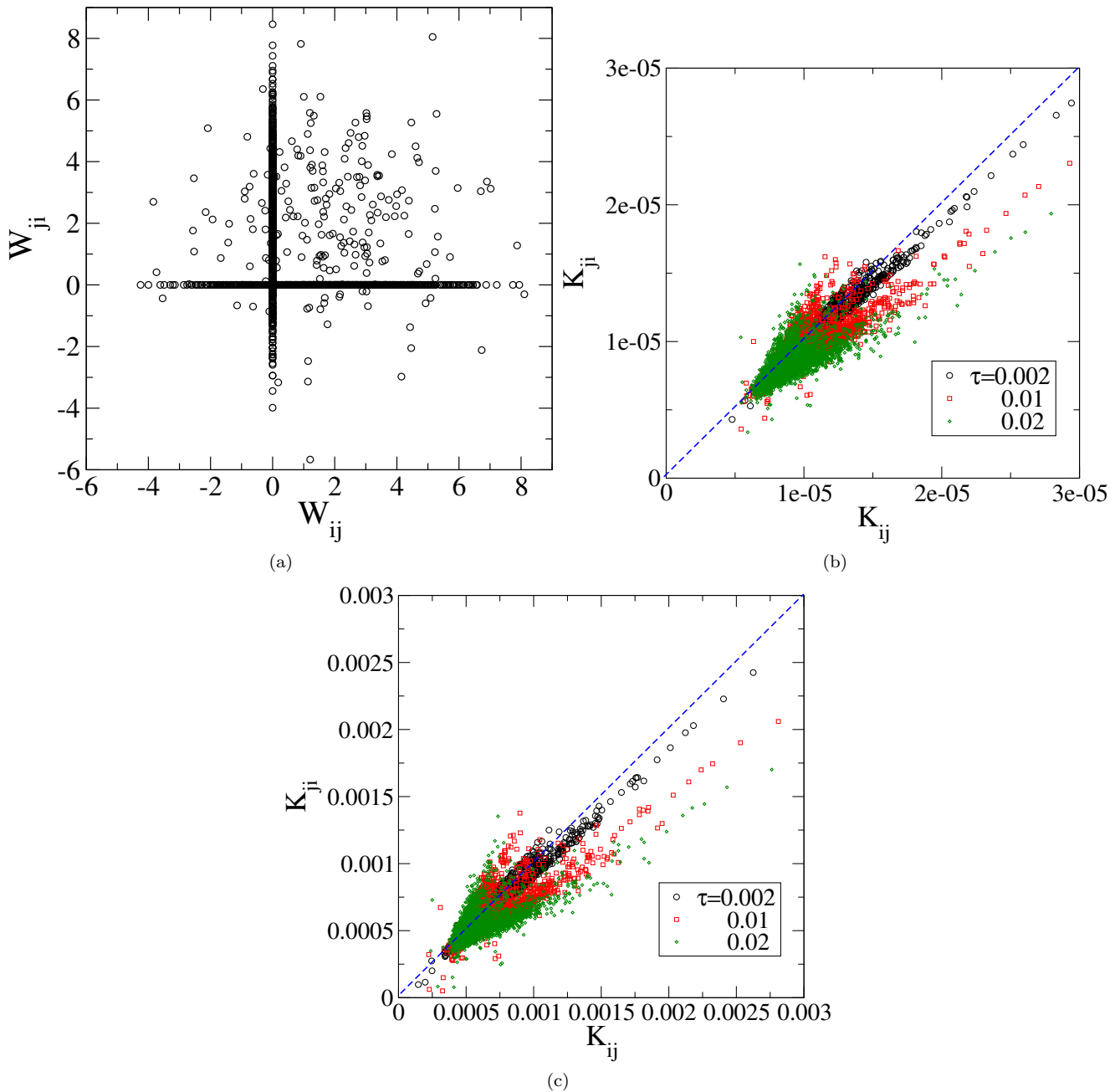

    \centering
  \subfigure[]{\includegraphics*[width=.48\columnwidth]{Fig9a.eps}}
  \subfigure[]{\includegraphics*[width=.48\columnwidth]{Fig9b.eps}}
  \subfigure[]{\includegraphics*[width=.48\columnwidth]{Fig9c.eps}}
    \caption{(a) Directed network with weights drawn from  a 
    mixture of two Gaussian distributions with means $\mu_1= 2$ and  $\mu_2= 4$ (with relative fraction of $\alpha= 0.2$
and $1-\alpha= 0.8$ respectively, with both standard deviations =2.  (b) Elements of the time-lag correlation matrix measured in simulation plotted against its transpose elements. The noise matrix is diagonal and uniform, with $\sigma_{ii}=0.01$. (c) Similar to (b) but with a stronger noise strength of $\sigma_{ii}=1$ . }
    \label{KijDWR} 
\end{figure}

Now we consider a more general Brownian system with forces that contain non-conservative parts, then one has
\begin{equation}
-\boldsymbol{\zeta} \dot{\vec{x}} + \vec{\mathscr{F}}(\vec{x}) + \vec{F}_r(t) = 0; \quad \nabla \vec{\mathscr{F}} \text{ is not symmetric}\label{LangF}
\end{equation}
with the noise variance still given by (\ref{FrFr}) describing the thermal bombardment from liquid molecules (and thus the
reciprocal relation is automatic: $\boldsymbol{\zeta} =\boldsymbol{\zeta} ^\intercal$). Here stable mechanical equilibrium is assumed $\vec{\mathscr{F}}(\vec{X})=0$, and ${\bf M}\equiv \nabla \vec{\mathscr{F}}|_{\vec{X}}$
has all eigenvalues with negative real parts. 
 (\ref{LangF}) is of the form
of network dynamics described in (\ref{xidot}) whose linearized dynamics 
 is given by (\ref{Deltaxdot}) with ${\bf Q}=\boldsymbol{\zeta}^{-1}{\bf M}$ and $\boldsymbol{\sigma} $ given by (\ref{sigzeta}). Then ${\bf Q}\boldsymbol{\sigma} =2k_B T\boldsymbol{\zeta}^{-1} {\bf M} \boldsymbol{\zeta}^{-1} $. Since $ {\bf M}$ is generally asymmetric, thus  ${\bf Q}\boldsymbol{\sigma}$ is also asymmetric and hence exemplifying the general non-equilibrium nature of a Brownian system under a non-conservative force.
In this case ${\bf C}_\tau\neq {\bf C}_\tau^\intercal$ and  ${\bf K}_\tau\neq {\bf K}_\tau^\intercal$.
 Fig. \ref{KijDWR}  shows the simulation results of a directed weighted random network with asymmetric weights (see Fig. \ref{KijDWR}a). The measured time-lag correlation function matrix elements are plotted against the elements of its transpose in Fig. \ref{KijDWR}b and \ref{KijDWR}c, verifying the the time-irreversibility nature of ${\bf K}_\tau\neq {\bf K}_\tau^\intercal$ in the NESS network dynamics. 

Using the  generalized FDR (\ref{FDR}), one obtains a generalized NESS FDR for
Brownian motion under a non-conservative force:
\begin{equation}
{\bf M} {\bf K}_0 \boldsymbol{\zeta} +  \boldsymbol{\zeta} {\bf K}_0 {\bf M}^\intercal + 2k_B T  \boldsymbol{\zeta} = 0.\label{nonconserFDR}
\end{equation}
If ${\bf M}$ has no asymmetric part (pure conservative forces), then 
${\bf MK}_0=-k_BT{\bf I}$ and reduces back to the equilibrium
case as before.

It is worth noting that even though $\vec{\mathscr{F}}$ is not derivable from a potential, one may construct an effective potential from ${\bf M}$ and $ \boldsymbol{\zeta}$. From the
solution of the Lyapunov equation (\ref{nonconserFDR}), one has
\begin{equation}
{\bf K}_0 = 2k_B T \int_0^{\infty} dt \, e^{t \ \boldsymbol{\zeta}^{-1} {\bf M}} \boldsymbol{\zeta}^{-1} e^{t {\bf M}^\intercal \boldsymbol{\zeta}^{-1}}.\label{ckmumu}
\end{equation}
 In the presence of non-conservative time-independent forces, one has a NESS\cite{PNAS2005,JDNoh2015} and the steady-state distribution $\psi_{ss}(\vec{x})$ is
expected to be of Gaussian form with $\psi_{ss}$  having 
the same form as in Eq. (\ref{psiss}) with the effective potential  given by $\Phi=\frac{1}{2}\delta{\vec x}^\intercal {\bf K}_0 ^{-1}\delta{\vec x} $.  Thus the  Brownian system under a non-conservative force can be described by an effective potential given by
\begin{equation}
\Phi = \frac{1}{4k_BT} \delta \vec{x}^\intercal \left( \int_0^{\infty} dt \, e^{t  \boldsymbol{\zeta}^{-1} {\bf M}}
\boldsymbol{\zeta}^{-1} e^{t {\bf M}^\intercal \boldsymbol{\zeta}^{-1}} \right)^{-1} \delta \vec{x}.
\end{equation}
For the case of $[\boldsymbol{\zeta},{\bf M}]= 0$, (and hence 
$[\boldsymbol{\zeta},{\bf M}^\intercal]=[\boldsymbol{\zeta}^{-1},{\bf M}]=[\boldsymbol{\zeta}, \boldsymbol{\zeta}^{-1}{\bf M}]=
[\boldsymbol{\zeta}^{-1}{\bf M}, {\bf M}^\intercal\boldsymbol{\zeta}^{-1}]=[{\bf M},{\bf M}^\intercal]=0$), one has
\begin{equation}
\Phi= 
\frac{1}{2k_BT} \delta \vec{x}^\intercal \left( \frac{{\bf M} + {\bf M}^\intercal}{2} \right) \delta \vec{x}\label{PhiMM}
\end{equation}
in the linearized regime.

For the special case of $\boldsymbol{\zeta}\propto {\bf I}$, as in the NESS of an overdamped Brownian colloid under a linear force, \eqref{PhiMM} agrees with the result in \cite{JDNoh2015}.
The NESS distribution can then be calculated from $\psi_{ss}(\vec x)\sim e^{-\Phi(\vec x)} $. Since the high-dimensional distribution $\psi_{ss}(\vec x)$ is not convenient to measure, one can instead measure the projected distribution of each node $\Psi_i(x_i)$, as given by 
\begin{equation}
    \Psi_i( {x_i})=\int\prod_{j\neq i} ^N d  {x_i} \psi_{ss}(\vec x).\label{Psixi}
\end{equation}
Fig. \ref{logPsi2}c show the reduced distribution $\Psi_i$ for a directed network with uniform diagonal noises, verifying its Gaussian nature as predicted. The widths of  ten $\Psi_i(x_i)$ are also obtained by fitting and compared against with the theoretical values given by \eqref{Psiixi} as displayed in Fig. \ref{logPsi2}d, showing perfect agreement.

\section{FLUCTUATION-DISSIPATION RELATION FOR NETWORK DYNAMICS under noise}
For general network dynamics under Gaussian white noises,  the equation can be written in the following suggestive form
\begin{equation}
-\boldsymbol{\zeta} \dot{\vec{x}} + \vec{\mathscr{F}}(\vec{x}) + \vec{ F}_r(t) = 0; \quad \nabla \vec{\mathscr{F}} \text{ is not symmetric in general}.\label{LangNet}
\end{equation}
Note that in general, the noise covariance of $ \vec{ F}_r$ is not related to $\boldsymbol{\zeta} $ via the usual FDR (\ref{FrFr}), and reciprocal
relation for $\boldsymbol{\zeta} $ does not hold  (i.e. $\boldsymbol{\zeta}\neq \boldsymbol{\zeta}^\intercal $). In this case, ${\bf C}_\tau\neq {\bf C}_\tau^\intercal$ and  ${\bf K}_\tau\neq {\bf K}_\tau^\intercal$ (numerical simulations verify this in Fig. 5 for the directed WR network).  For the network system (\ref{LangNet}) that has a noise-free stable state at $\vec{X}$ , it follows
from (\ref{FDR}) that the generalized FDR for the general dynamics governed by (\ref{LangNet}) (Onsager reciprocal relation may not hold) reads
\begin{eqnarray}
\langle \vec{F}_r(t) \vec{F}_r^\intercal (t') \rangle &=&-({\bf M} {\bf K}_0\boldsymbol{\zeta}^\intercal + \boldsymbol{\zeta} {\bf K}_0 {\bf M}^\intercal) \delta(t - t')\label{FDRFrFr}
\end{eqnarray}
where stable mechanical equilibrium is assumed with $\vec{\mathscr{F}}(\vec{X})=0$, and ${\bf M}\equiv \nabla \vec{\mathscr{F}}|_{\vec{X}}$ whose eigenvalues have no positive real parts. One expects the steady-state distribution $\psi_{ss}(\vec{x})$ still obeys (\ref{psissPhi}) with a Gaussian form.
And for the equilibrium case that obeys (\ref{KQQK}), \eqref{FDRFrFr} reduces to  $\langle \vec{F}_r(t) \vec{F}_r^\intercal (t') \rangle =-2\boldsymbol{\zeta}  {\bf K}_0 {\bf M}^\intercal\delta(t - t')$.

Now we examine what types of networks governed by (\ref{Deltaxdot}) can be viewed as a Brownian system fluctuating at equilibrium obeying FDT and the Boltzmann distribution. In this case, one needs to construct a potential $U$ for conservative forces which is possible
under suitable conditions. Following discussions in previous sections, it follows that if ${\bf Q}\boldsymbol{\sigma}$  is symmetric, then the
noisy network dynamics is a Brownian  type at equilibrium. From (\ref{KKtrans})-(\ref{QsigsigQ}), and assuming all matrix inverses exist, a symmetric
${\bf Q}\boldsymbol{\sigma}$ implies $\boldsymbol{\sigma}^{-1}{\bf Q}$ is also symmetric and can be identified as the Hessian matrix $-\frac{{\bf H}}{2k_BT}$. 
Furthermore, Boltzmann distribution holds at
equilibrium (as verified by ${\bf K}_\tau= {\bf K}_\tau^\intercal$  in Fig. \ref{Qsigsymm}d) with
\begin{equation}
\psi_{\text{eqm}}(\vec{x}) = \frac{e^{\Delta \vec{x}^\intercal \boldsymbol{\sigma}^{-1} {\bf Q} \Delta \vec{x}}}{\pi^{N/2} \sqrt{\det {\bf Q}^{-1} \det \boldsymbol{\sigma}}}\sim e^{-\frac{\Delta \vec{x}^\intercal{\bf H}\Delta \vec{x}}{2k_B T}}.
\end{equation}

\section{Conclusion and Outlook}
In this paper, we  investigated the fluctuating dynamics of general noisy networks about a stable steady state. The key assumption is that the noise-free dynamics has a stable fixed point resulting in steady-state equilibrium or nonequilibrium fluctuating dynamics. Connections between the  traditional theory of physical Brownian motion and the broader class of stochastic dynamics on directed, nonlinear networks are examined and established.
Our key contributions are multi-fold. First, we derive several equivalent conditions for equilibrium in terms of the symmetry of time-lag correlation matrices and the commutation relation ${\bf Q} \boldsymbol{\sigma}$ = $\boldsymbol{\sigma} {\bf Q}^\intercal$. This clarifies how non-uniform noise or asymmetric connections independently break detailed balance. Second, we demonstrate that conventional overdamped Brownian dynamics is a special case of our general framework, occurring only under restrictive conditions on ${\bf Q}$ and $\boldsymbol{\sigma}$. Third, we extend the concept of the Fluctuation-Dissipation Relation (FDR) to general NESSs, providing a formula (Eq. \eqref{FDR}) that holds universally for both equilibrium and non-equilibrium steady-state dynamics. We also analyzed the decomposition of the dynamical force into dissipative and non-dissipative components, linking the antisymmetric part of ${\bf QK}_0$ directly to the steady-state probability current on the equi-$\Phi$ surface. Finally, we connect our theoretical results to practical applications on network reconstruction from time-series data, emphasizing on how the measured correlation matrix ${\bf K}_\tau$ can be used to infer the underlying connection weights ${\bf W}$ and noise strengths $\boldsymbol{\sigma}$.

The theoretical findings are verified by numerical simulations of nonlinear network dynamics on Erdős–Rényi-type networks, considering both bidirectional and directed topologies with heterogeneous noise. This work not only deepens the theoretical understanding of non-equilibrium statistical physics in networked systems but also provides deeper insights on the methods of reconstructing complex systems from noisy observational data. It establishes a foundation for future investigations into the thermodynamics of networks, including entropy production and energy flows at the nodal level.

We further obtained generalized FDRs for NESS in noisy network dynamics (Eqs. \eqref{FDR} and \eqref{FDRFrFr}) and for physical Brownian systems (Eqs. \eqref{sigzeta} and \eqref{nonconserFDR}).
The study of fluctuation-dissipation relations in network dynamics reveals the complex interplay between noise structure and network connectivity, where network reconstruction can inform dynamic properties, time-reversal symmetry, and out-of-equilibrium behavior. 
It should be noted that there are several studies on the generalization of FDR to NESS focussing on the correction due to the finite entropy production rate, house-keeping dissipation rate or non-vanishing currents\cite{Seifert2010,Mehl2010,Altaner2016,Aslyamov2025}.

Our present work examined on the connection between general network dynamics under noise and the conventional non-equilibrium steady-state dynamics in the context of statistical physics. We focused here on the simpler case of stable noise-free dynamics, and the noisy fluctuations around this stable fixed point give rise to the equilibrium or non-equilibrium steady states, which are governed by the connection and noise matrices. This relies on the assumption that the matrix ${\bf Q}$ is stable, i.e. the real parts of all its eigenvalues are negative. As shown in Appendix D, ${\bf Q}$ is stable if the intrinsic  node dynamics are stable near the fixed point (i.e. $f_i'(X_i)<0$) and the connection weights are non-negative ($W_{ij}\geqslant 0$).
  On the other hand, for general network systems whose noise-free dynamics are beyond the stable fixed point considered here, the resulting dynamics can depend on the network's structural properties in general. For instance, depending on the intrinsic dynamics ($f_i$), the connection matrix ${\bf W}$, and the coupling function $h(x_i,x_j)$, the  asymptotic dynamics\cite{chengLaiurns2023} can lie on a stable limit cycle (periodic) or even chaotic (such as a strange attractor). 

Our results derived in this work is limited within the linearized framework. Presumably, nonlinear effects in noisy network dynamics can be examined systematically via perturbations, using similar methods as in Ref. \cite{KwonAo2011}. In addition, higher-order nonlinear effects can lead to multiple (coexisting) noise-free stable fixed points, and give rise to the non-Gaussian behavior in the steady-state distribution. It would be interesting to investigate systems with multiple and distinct steady states and examine the corresponding equilibrium or non-equilibrium phase transitions between these states from the viewpoints of network dynamics and stochastic thermodynamics. 

There are several related topics that can be further investigated. First, by viewing the network dynamics as an overdamped Langevin system, one can further identify the associated energetics, such as heat, work, and entropy production rates for general linearized noisy networks at NESSs, one can then theoretically derive the network energetics in terms of the network structure properties and other relevant measurable quantities such as the time correlation functions. The network energetics associated with each node and their dependence on the network properties are under current investigation.
Second, for
network dynamics fluctuates around the noise-free  periodic state\cite{Lai2017}, the associated non-equilibrium periodic states\cite{chengLaiurns2023} can be systematically investigated. Third, for the interesting case of
 network dynamics under temporally correlated (colored) noises\cite{TCL2018},  a deeper understanding on possible fluctuation-dissipation relations  can provide deeper insight on the overdamped Brownian systems under active reservoirs\cite{Wu2000,Albay2023,Bebon2025,Goswami2025,Goswami2026}.
 Finally it would be of interesting to compare our results with  time-series data in some empirical realistic network systems, such as steady-state fluctuating expression dynamics in protein or genetic networks.

\appendix
   \section{Derivations of Eqs. (\ref{Ctau}) and (\ref{QCCQ}) }
  
  Eq. \eqref{Ctau} was derived in Ref. \cite{CT2017}, and  the derivation is shown here again for completeness.
  The steady-state solution of the linearized motion in (\ref{Deltaxdot}) is (see \eqref{Deltaxt})
\begin{equation}
    \Delta{\vec x}(t)=\int^t_0 dt' e^{ (t-t'){\bf Q}}{\vec \eta}(t').\label{Deltaxtss}
\end{equation}
Then the time-lag correlation function ${\bf C}_\tau$ is (without loss of generality, take $\tau\geq 0$)
\begin{eqnarray}
 {\bf C}_\tau&\equiv &\langle\Delta{\vec x}(t+\tau) \Delta{\vec x}^\intercal(t)\rangle=\int^{t+\tau}_0 dt'\int^t_0 dt''  e^{ (t+\tau-t'){\bf Q}}\langle{\vec \eta}(t'){\vec \eta}^\intercal(t'')\rangle e^{ (t-t''){\bf Q}^\intercal}\\
 &=&e^{\tau{\bf Q}}\int^{t+\tau}_0 dt'\int^t_0 dt''  e^{ (t-t'){\bf Q}}\boldsymbol{\sigma}\delta(t'-t'') e^{ (t-t''){\bf Q}^\intercal}\\
 &=&e^{\tau{\bf Q}}\int^{t}_0 dt'\int^t_0 dt''  e^{ (t-t'){\bf Q}}\boldsymbol{\sigma}\delta(t'-t'') e^{ (t-t''){\bf Q}^\intercal}\\
  &=&e^{\tau{\bf Q}} {\bf C}_0.
\end{eqnarray}

Using the expression of  ${\bf C}_0=\int^{t}_0 dt'\int^t_0 dt''  e^{ (t-t'){\bf Q}}\boldsymbol{\sigma}\delta(t'-t'') e^{ (t-t''){\bf Q}^\intercal}$, integrating over the $\delta$-function, we have
\begin{eqnarray}
 {\bf Q}{\bf C}_0+{\bf C}_0 {\bf Q}^\intercal &=&\frac{1}{2}
 \int^{t}_0 dt'  e^{ (t-t'){\bf Q}}({\bf Q}\boldsymbol{\sigma}+\boldsymbol{\sigma}{\bf Q}^\intercal)  e^{ (t-t'){\bf Q}^\intercal}\\
 &=& \frac{{\bf Q}}{2}\left(\int^{t}_0 dt'  e^{ (t-t'){\bf Q}}\right)\boldsymbol{\sigma}+\boldsymbol{\sigma} \frac{{\bf Q}^\intercal}{2} \left(\int^{t}_0 dt'  e^{ (t-t'){\bf Q}^\intercal}\right)
  \\&=&\frac{{\bf Q}}{2}(-{{\bf Q}^{-1}})\boldsymbol{\sigma}+\boldsymbol{\sigma} \frac{{\bf Q}^\intercal}{2}(-{\bf Q}^{-\intercal})
 \\&=&-\boldsymbol{\sigma},
\end{eqnarray}
where the identity $\int^{t}_0 dt'  e^{ (t-t'){\bf Q}}={\bf Q}^{-1}(e^{ t{\bf Q}}-{\bf I}) \xrightarrow{t\to\infty}-{\bf Q}^{-1}$ is used.
  \section{Solution of the Lyapunov Equation by Spectral decomposition}
     Consider the Lyapunov Equation for the unknown ${\bf K}$ of the form
 \begin{equation}
  {\bf Q} {\bf K}+ {\bf K} {\bf Q}^\intercal={\bf S}\label{lyap}
  \end{equation}    
  where $ {\bf Q}$ and ${\bf S}$ are given matrices and ${\bf S}$ is symmetric.
Assuming ${\bf Q}$ can be diagonalized in terms of its eigenvalues $\lambda_1,\lambda_2,\cdots,\lambda_N$ and  eigenvectors ${\vec u_1},{\vec u_2},\cdots,{\vec u_N}$, i.e. ${\bf Q}{\vec u_i}=\lambda_i{\vec u_i}$ and ${\bf P}^{-1}{\bf Q}{\bf P}=\boldsymbol{\Lambda}\equiv\hbox{diag}(\lambda_1,\lambda_2,\cdots,\lambda_N)$, where ${\bf P}=({\vec u_1},{\vec u_2},\cdots,{\vec u_N})$ is the  matrix of the eigenvectors.
Then by defining ${\tilde {\bf K}}={\bf P}^{-1}{\bf K}{\bf P}^{-\intercal}$, (\ref{lyap})  gives
\begin{eqnarray}
& &  \boldsymbol{\Lambda}{\tilde {\bf K}}+ {\tilde {\bf K}}\boldsymbol{\Lambda}  ={\bf P}^{-1}{\bf S}{\bf P}^{-\intercal}\\
&\implies& ({\tilde {\bf K}})_{ij}=\frac{({\bf P}^{-1}{\bf S}{\bf P}^{-\intercal})_{ij}}{\lambda_i+\lambda_j}.\label{Ktil}
\end{eqnarray}
Hence, the solution of the Lyapunov Eq. (\ref{lyap}) is
\begin{equation}
 {\bf K}={\bf P}  {\tilde {\bf K}}{\bf P}^\intercal
\end{equation}
where ${\tilde   {\bf K}}$ is given by (\ref{Ktil}).
Note that for a noisy network that fluctuates about some stable noise-free solution, the real part of all the eigenvalues are negative, and hence $\lambda_i+\lambda_j\neq 0$ in (\ref{Ktil}).

The above solution can also be written in terms of the spectral decomposition of ${\bf Q}$ with
\begin{equation}
    {\bf Q}=\sum_{\ell=1}^N \lambda_\ell {\bf M}^{(\ell)},
\end{equation}
where the matrices ${\bf M}^{(\ell)}$ formed by   is  given  by 
\begin{equation}
{\bf M}^{(\ell)}\equiv  {\vec u_\ell} {\vv v_\ell}^\intercal ,\quad \ell=1,2,\cdots,N  
\end{equation}
where $ {\vec v_\ell}$ are the vectors forming ${\bf P}^{-\intercal}$: ${\bf P}^{-\intercal}=({\vec v_1},{\vec v_2},\cdots,{\vec v_N})$ and $ {\vv v_i}^\intercal  {\vec u_j}=\delta_{ij}$, $\sum_{j=1}^N{\vec v_j} {\vv u_j}^\intercal={\bf I}$.
In terms of ${\bf P}$ and its inverse, the elements of ${\bf M}$ are given by $
    M_{ij}^{(\ell)}=P_{i\ell}P^{-1}_{\ell j}$.
Then the solution of $ {\bf K}$  is given by
\begin{equation}
 {\bf K}=\sum_{\ell,m=1}^N \frac{ {\bf M}^{(\ell)}{\bf S} {{\bf M}^{(m)}}^\intercal }{\lambda_\ell+\lambda_m}.
\end{equation}

For the general case of possible degenerate eigenvalues or ${\bf Q}$ can have unstable eigenvalues such that $\lambda_\ell+\lambda_m $ might be zero, one can employ the Jordan transformation of a real matrix to handle such singular cases\cite{PNAS2005}.

\section{Derivations of the equivalent conditions from (\ref{KKtrans}) to (\ref{KQQK}) for equilibrium Network dynamics}
We first show that under steady-state conditions (i.e. time-translational invariant) a time-reversal symmetric time-lag correlation matrix
 equivalent to a symmetric  ${\bf K}_\tau$ matrix for all $\tau$:
 \begin{equation}
    {\bf K}_\tau= {\bf K}_{-\tau}\Leftrightarrow  \langle \delta{ x}_i(\tau)\delta{ x}_j(0)\rangle = \langle \delta{ x}_j(0)\delta{ x}_i(\tau)\rangle = \langle \delta{ x}_j(\tau)\delta{ x}_i(0)\rangle 
  \Leftrightarrow {\bf K}_\tau= {\bf K}_\tau^\intercal\end{equation} 
  where time-translational invariance is used in the second equality above.  For general nonlinear $\vec{\cal F}(\vec x)$, $\vec X=\vec x_{peak}\neq \langle\vec x\rangle$ and $\langle\vec x\rangle=\vec X+\vec \mu$ even at equilibrium.
Since the correlation function $ {\bf C}_\tau\equiv \langle\Delta{\vec x}(\tau) \Delta{\vec x}^\intercal(0)\rangle$ is related to $ {\bf K}_\tau$ via (\ref{KCmu}), it follows that
  \begin{equation}
      {\bf K}_\tau= {\bf K}_\tau^\intercal \Leftrightarrow  {\bf C}_\tau-\vec\mu \vec\mu^\intercal= {\bf C}_\tau^\intercal-(\vec\mu \vec\mu^\intercal)^\intercal\Leftrightarrow  {\bf C}_\tau= {\bf C}_\tau^\intercal.\label{KtauCtaueqm}
  \end{equation}
  Notice that the equilibrium conditions \eqref{KtauCtaueqm} holds in general, even beyond the linearized regime.
  
 We can then consider the  equilibrium fluctuating dynamics in the linearized regime. Since there is no distinction between $ {\bf K}_\tau $ and ${\bf C}_\tau$ in the linearized regime, here we prove the equilibrium conditions in terms of $ {\bf C}_\tau$.
    We first show that if ${\bf QC}_0= {\bf C}_0{\bf Q}^\intercal$, then
  \begin{eqnarray}
    {\bf C}_\tau= e^{\tau{\bf Q}}{\bf C}_0&=&\sum_{n=0}^\infty \frac{\tau^n}{n!}{\bf Q}^n{\bf C}_0\\
    &=&\sum_{n=0}^\infty \frac{\tau^n}{n!}{\bf C}_0({\bf Q}^\intercal)^n={\bf C}_0 e^{\tau{\bf Q}^\intercal}= {\bf C}_\tau^\intercal.
      \end{eqnarray}
      Conversely,  
    \begin{eqnarray}
    {\bf C}_\tau= {\bf C}_\tau^\intercal &\implies& e^{\tau{\bf Q}}{\bf C}_0= {\bf C}_0 e^{\tau{\bf Q}^\intercal}\implies\sum_{n=0}^\infty \frac{\tau^n}{n!}\left({\bf Q}^n{\bf C}_0-{\bf C}_0({\bf Q}^\intercal)^n\right)=0\quad\hbox{for arbitrary } \tau\\
    &\implies&{\bf Q}^n{\bf C}_0= {\bf C}_0({\bf Q}^\intercal)^n\quad\hbox{for } n=0,1,2,\cdots\label{QKKQApp}
  \end{eqnarray}
Hence ${\bf QC}_0= {\bf C}_0{\bf Q}^\intercal \Leftrightarrow{\bf C}_\tau= {\bf C}_{\tau}^\intercal $.

 If ${\bf QC}_0= {\bf C}_0{\bf Q}^\intercal$, then from (\ref{FDR}), one has
  \begin{eqnarray}
    & &  {\bf C}_0{\bf Q}^\intercal=-\frac{\boldsymbol{\sigma}}{2} \implies 
      \boldsymbol{\sigma} {\bf Q}^\intercal=-\frac{\boldsymbol{\sigma} {\bf C}_0^{-1}\boldsymbol{\sigma} }{2}\\
 & &    {\bf Q} {\bf C}_0=-\frac{\boldsymbol{\sigma}}{2} \implies 
       {\bf Q}\boldsymbol{\sigma}=-\frac{\boldsymbol{\sigma} {\bf C}_0^{-1}\boldsymbol{\sigma} }{2}\\
       &\implies&  {\bf Q}\boldsymbol{\sigma}=\boldsymbol{\sigma} {\bf Q}^\intercal, \hbox{ i.e. } {\bf Q}\boldsymbol{\sigma} \hbox{ is symmetric. }\label{QsigsigQApp}
  \end{eqnarray}
  Conversely, starting from ${\bf Q}\boldsymbol{\sigma}=\boldsymbol{\sigma} {\bf Q}^\intercal$ and using the formal solution of the Lyapunov equation in (\ref{Lynsoln}), one has
  \begin{eqnarray}
      {\bf Q} {\bf C}_0&=&\int_0^\infty dt e^{t{\bf Q}} {\bf Q}\boldsymbol{\sigma}e^{t{\bf Q}^\intercal}=\int_0^\infty dt e^{t{\bf Q}}\boldsymbol{\sigma}{\bf Q}^\intercal e^{t{\bf Q}^\intercal}\\
      &=&\left(\int_0^\infty dt e^{t{\bf Q}}\boldsymbol{\sigma}e^{t{\bf Q}^\intercal}  \right){\bf Q}^\intercal ={\bf C}_0 {\bf Q}^\intercal.\label{QCoCoQ}
        \end{eqnarray}
          Hence the equivalent equilibrium conditions: $ {\bf C}_\tau= {\bf C}_\tau^\intercal \Leftrightarrow{\bf QC}_0= {\bf C}_0{\bf Q}^\intercal =-\frac{\boldsymbol{\sigma}}{2}\Leftrightarrow {\bf Q}\boldsymbol{\sigma}=\boldsymbol{\sigma} {\bf Q}^\intercal$ are proved,  and the conditions (\ref{KKtrans}) to (\ref{KQQK}) hold under equilibrium.

\begin{figure}[H]
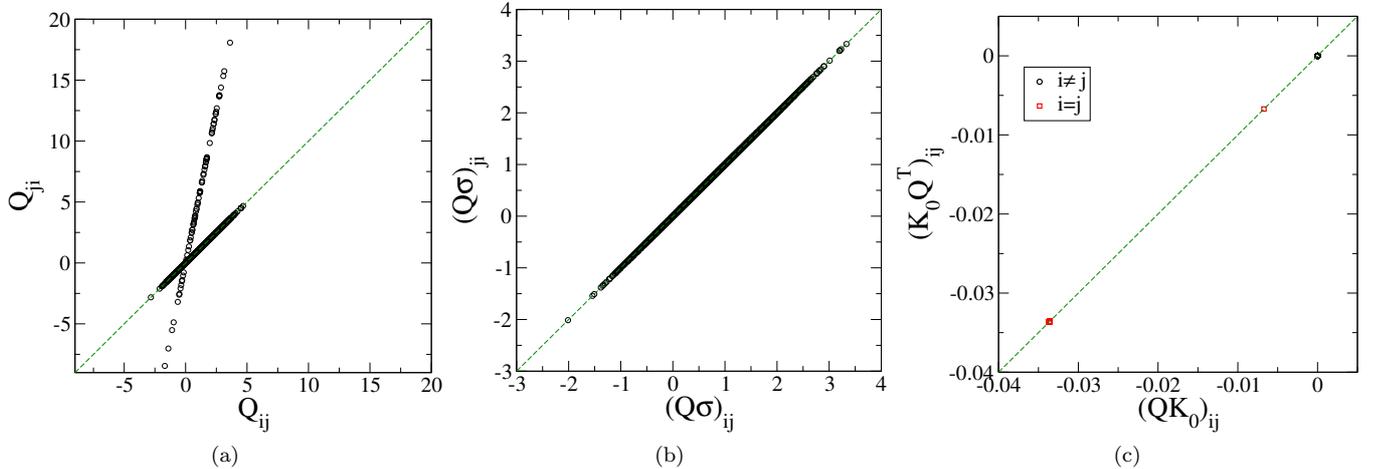

    \centering
 \subfigure[]{\includegraphics*[width=.32\columnwidth]{Fig10a.eps}}
  \subfigure[]{\includegraphics*[width=.32\columnwidth]{Fig10b.eps}}
 \subfigure[]{\includegraphics*[width=.34\columnwidth]{Fig10c.eps}}
          \caption{Network with $N=100$ nodes with asymmetric ${\bf Q}$, but non-trivial diagonal noise matrix adjusted so that  ${\bf Q}\boldsymbol{\sigma}$ is still symmetric and equilibrium is achieved. The noises are non-uniform with $\sigma_{ii}$ taking two different values.  ${\bf C}_0$ is obtained from the solution of the Lyapunov equation \eqref{QCCQ} using the method described in Appendix B. (a) $Q_{ji}$ vs. $Q_{ij}$ showing the asymmetric ${\bf Q}$. (b) Off-diagonal elements of ${\bf Q}\boldsymbol{\sigma}$ vs. its transpose
showing the noise is constructed to make ${\bf Q}\boldsymbol{\sigma}$ symmetric. (c) Elements of ${\bf C}_0{\bf Q}^\intercal$ vs. ${\bf QC}_0$  verifying (\ref{QKKQApp}).}\label{QKLyap}
    \end{figure}

To verify the equilibrium conditions given by Eq. (\ref{QKKQApp}) for $n=1$ and Eq. (\ref{QCoCoQ}), we prepare a network with asymmetry in ${\bf Q}$ (Fig. \ref{QKLyap}a), but with a
diagonal $\boldsymbol{\sigma}$ adjusted such that the matrix ${\bf Q}\boldsymbol{\sigma}$ is symmetric (Fig. \ref{QKLyap}b). Then  ${\bf C}_0$ is solved from the Lyapunov equation \eqref{QCCQ} using the method described in Appendix B.  The elements of    ${\bf C}_0{\bf Q}^\intercal$ are plotted against the elements of ${\bf QC}_0$    in Fig. \ref{QKLyap}c, verifying ${\bf C}_0{\bf Q}^\intercal={\bf QC}_0$. Note that since $\boldsymbol{\sigma}$ is chosen to be diagonal and $\sigma_{ii}$ take only two distinct values in this example, thus the diagonal elements of both ${\bf C}_0{\bf Q}^\intercal$ and $={\bf QC}_0$ are zero, and that the their off-diagonal elements take only two distinct values.
    
    \section{ Stability analysis of the matrix ${\bf Q}$}
In this paper, it is assumed that the linearized dynamics is  stable, i.e. the real parts of all the eigenvalues of ${\bf Q}$ (Jacobian matrix at the noise-free fixed point) are negative. Here we examine the conditions, in terms of the network  properties, for such an assumption to hold. Here we consider the mutual synchronizing coupling function between two nodes  of the form $h(x_i,x_j)=\tilde{h}(x_j-x_i)$, with $\tilde{h'}_{ij}\equiv\tilde{h}'(X_j-X_i)>0$. Then the matrix elements of ${\bf Q}$ can be obtained from \eqref{Qij} (with  $\tilde{W}_{ij}\equiv W_{ij}\tilde{h'}_{ij}$) to give
\begin{equation}
 Q_{ij}= \tilde{W}_{ij}+\left[  f_i'(X_i;r_i)- \tilde{ \kappa}_i \right]\delta_{ij}.\label{Qij2}
\end{equation}
where $\tilde{ \kappa}_i\equiv \sum_m\tilde{W}_{im}$ is the scaled weighted in-degree of node $i$. Since $\tilde{h'}_{ij}>0$, the $\tilde{}$ on $W_{ij}$ can be dropped hereafter for notation simplicity.

In this Appendix, we will show that for networks with $W_{ij}\geqslant 0$ and stable intrinsic node dynamics, $f_i'\equiv f_i'(X_i;r_i)<0$, 
the real parts of all the eigenvalues of ${\bf Q}$ are negative. Note that the diagonal elements of ${\bf Q}$, $Q_{ii}=f_i'-\kappa_i$, is always negative. We will first prove explicitly that that above claim is true for $N=2$ and $N=3$.
For $N=2$, ${\bf Q}= \begin{pmatrix}
f_1-W_{12} & W_{12} \\
W_{21} & f_2'-W_{21} 
\end{pmatrix}$, and direct calculation reveals that the stability condition for ${\bf Q}$ is $(f'_1-W_{12})(f_2'-W_{21} )> W_{12} W_{21} $ or $ \frac{W_{12} }{f'_1}+\frac{W_{21} }{f'_2}<1$. It is then clear that the stable condition is satisfied for $f'_i<0$ and $W_{ij}\geqslant 0$.

For $N=3$,
\begin{equation}
{\bf Q}= \begin{pmatrix}
f'_1-W_{12}-W_{13} & W_{12} & W_{13}\\
W_{21} & f_2'-W_{21}-W_{23} & W_{23}\\
W_{31}  & W_{32}& f_3'-W_{31}-W_{32}
\end{pmatrix}.
\end{equation}
It is easy to show by direct calculations that the eigenvalues are given by the roots of the characteristic polynomial $P(\lambda)=\lambda^3- \text{Tr}{\bf Q} \lambda^2+\text{Tr}(\text{adj}( {\bf Q}) )\lambda-\text{det} {\bf Q}$, where adj$({\bf Q})$ denotes the adjoint of ${\bf Q}$.  Direct calculation gives
\begin{eqnarray}
\text{Tr}{\bf Q}&=&f'_1+f'_2+f'_3-W_{12}-W_{13}-W_{21}-W_{23}-W_{31}-W_{32} \\
\text{det} {\bf Q}&=&f'_1f'_2f'_3-(W_{31}+W_{32})f'_1f'_2
-(W_{12}+W_{13})f'_2f'_3-(W_{21}+W_{23})f'_3f'_1\\
& & +(W_{21}W_{31}+W_{21}W_{32}+W_{23}W_{31} )f'_1\nonumber\\
& & +(W_{12}W_{31}+W_{12}W_{32}+W_{13}W_{32} )f'_2\nonumber\\
& & +(W_{12}W_{23}+W_{13}W_{23}+W_{13}W_{21} )f'_3.\nonumber
\end{eqnarray}
Since  $f'_i<0$ and $W_{ij}\geqslant 0$, it is clear that both $\text{Tr}{\bf Q}$ and det${\bf Q}$ are $<0 $.
According to the Routh-Hurwitz stability criterion, the eigenvalues of ${\bf Q}$ have negative real parts if and only if (i) both $\text{Tr}{\bf Q}$ and det${\bf Q}$ are negative, (ii) det${\bf Q}- \text{Tr}{\bf Q}\text{Tr}(\text{adj}( {\bf Q}) )>0 $. It remains to show that condition (ii) is satisfied. After some tedious calculations, we obtained
\begin{eqnarray}
\text{det}{\bf Q}- \text{Tr}{\bf Q}\text{Tr}(\text{adj}( {\bf Q}) )&=&(W_{12}+W_{13}+W_{21}+W_{23}+W_{31}+W_{32}) [W_{21}W_{31}+W_{21}W_{32}+W_{23}W_{31}\nonumber\\
& &+W_{12}W_{31}+W_{12}W_{32}+W_{13}W_{32}+W_{12}W_{23}+W_{13}W_{23}+W_{13}W_{21} \nonumber
\\& &+2(f_1'f_2'+f_2'f_3'+f_3'f_1')]+(W_{21}+W_{23}+W_{31}+W_{32})f_1'^2\nonumber
\\& &+(W_{12}+W_{13}+W_{31}+W_{32})f_2'^2+(W_{12}+W_{13}+W_{21}+W_{23})f_3'^2\nonumber\\& &
-[(W_{21}+W_{23}+W_{31}+W_{32})^2+ W_{12}(W_{21}+2W_{23}+2W_{31}+2W_{32})\nonumber\\& &
+W_{13}(2W_{21}+2W_{23}+W_{31}+2W_{32})]f_1'\nonumber\\& &
-[(W_{12}+W_{13}+W_{31}+W_{32})^2 + W_{21}(W_{12}+2W_{13}+2W_{31}+2W_{32})\nonumber\\& &+W_{23}(2W_{12}+2W_{13}+2W_{31}+W_{32})]f_2'\nonumber\\& &
-[(W_{12}+W_{13}+W_{21}+W_{23})^2+ W_{31}(2W_{12}+W_{13}+2W_{21}+2W_{23})\nonumber\\& &+W_{32}(2W_{12}+2W_{13}+2W_{21}+W_{23})]f_3'\nonumber\\& &
-(f_2'+f_3')f_1'^2-(f_1'+f_3')f_2'^2-(f_1'+f_2')f_3'^2-2f_1'f_2'f_3'.\label{RH}
\end{eqnarray}
Since   $f'_i<0$ and $W_{ij}\geqslant 0$ , it is clear that every term in \eqref{RH} is positive and hence condition (ii) also holds, proving that ${\bf Q}$ is stable for $N=3$.

For $N>3$, the algebra becomes too tedious for direct calculating the eigenvalues.
 However, for the special case of a fully connected network of $N$ nodes with equal weights ($W_{ij}=W\geqslant 0$) and identical intrinsic node dynamics ($f_i'=f'<0$), the eigenvalues of ${\bf Q}$ can be calculated to be: $\lambda =f'$ and $f'-NW$ ($(N-1)$ degeneracy), confirming that all eigenvalues are negative in this case.
For general $N$, one can invoke the Gershgorin circle theorem\cite{Gershgorin} which states that every eigenvalue of an arbitrary matrix (${\bf Q}$ here) lies within at least one of the $N$ Gershgorin discs. The center of the $i^{th}$ Gershgorin disc is at $Q_{ii}=f'_i-\kappa_i$ with a radius $R_i=\sum_{m\neq i}^N |Q_{im}|=\kappa_i$. Since $f'_i<0$, all Gershgorin discs must lie in the Re$\lambda<0$ regime of the complex plane of the eigenvalues and hence proving all real parts of the eigenvalues of ${\bf Q}$ must be negative for networks with  $f'_i<0$ and $W_{ij}\geqslant 0$.
 
\begin{figure}[H]
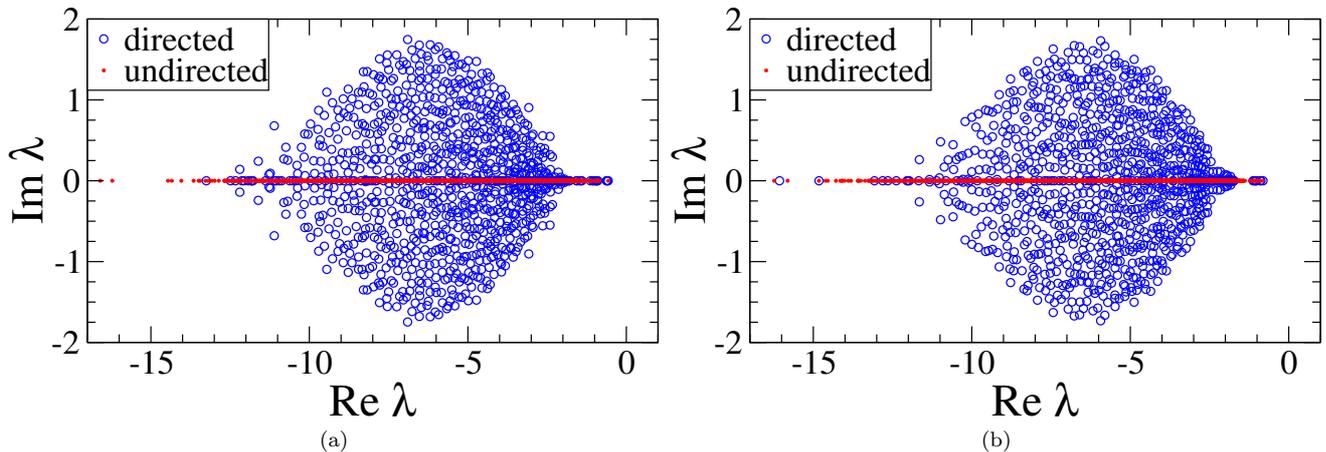

    \centering
  \subfigure[]{\includegraphics*[width=.48\columnwidth]{Fig11a.eps}}
  \subfigure[]{\includegraphics*[width=.48\columnwidth]{Fig11b.eps}}
    \caption{Eigenvalues of ${\bf Q}$ plotted in the complex plane for $N=1000$ nodes directed and undirected random networks with connection probability $p=0.005$. (a) $f_i'$ and $W_{ij}$ are Gaussian distributed with (mean, standard deviation) =(-1,0.2) and  (1,0.2) respectively. (b) $f_i'$ and $W_{ij}$ are uniformly distributed in [-1.2,-0.8] and [0.8,1.2] respectively. Note that for undirected network, ${\bf Q}$ is symmetric and its eigenvalues are all real. }
    \label{eigen}
    \end{figure}

To verify the proof, we  compute the eigenvalues of ${\bf Q}$ numerically for arbitrary networks with  $f'_i<0$ and $W_{ij}\geqslant 0$. In all the cases we tested, the the real parts of the eigenvalues of ${\bf Q}$ are negative. Fig. \ref{eigen} shows the plots of the eigenvalues in the complex plane for directed and undirected $N=1000$ networks with randomly distributed $f'_i$ and $W_{ij}$, demonstrating indeed that ${\bf Q}$ is stable for networks with   $f'_i<0$ and $W_{ij}\geqslant 0$. 

\section{Evaluate the distribution 	$\Psi(\delta x_i)$}
The projected distribution of each node $\Psi_i(x_i)$ is given by \eqref{Psixi} and \eqref{psiss}. Here we need to evaluate an integral of the form
\begin{equation}
\int \prod_{j \neq i} dx_j \, e^{-\frac{1}{2} \vec{x}^\intercal {\bf B} \vec{x}}\label{hiDint}
\end{equation}
for a real $N\times N$ symmetric matrix ${\bf B}$. Denote the $(N-1)\times (N-1)$ submatrix formed by deleting the $i$th row and
$i$th column by ${\cal B}_{\hat i}$ , and the corresponding deleted $(N-1)$-component column vector with $B_{ii}$ removed by $\vec{b}_i$. Let ${\bf P}$ be the orthogonal matrix that diagonalizes ${\cal B}_{\hat i}$ (i.e. ${\bf P}^\intercal  {\cal B}_{\hat i}{\bf P} = \text{diag}(\mu_1, \mu_2,\cdots \mu_{N-1})$), then by transforming $\vec{y} = {\bf P}^\intercal \vec{x}_{\hat i}$, the bilinear term in the exponent of the integrand in \eqref{hiDint} becomes
\begin{eqnarray}
  \vec{x}^\intercal {\bf B} \vec{x}&=&  \vec{x}_{\hat{i}}^\intercal {\bf B} \vec{x}_{\hat{i}} + 2 \sum_{j \neq i} B_{ji} x_i x_j + B_{ii} x_i^2\\
&=& \sum_{\alpha=1}^{N-1} \mu_{\alpha} y_{\alpha}^2 + 2 \sum_{\alpha=1}^{N-1} ({\bf P}^\intercal \vec{b}_{\hat{i}})_{\alpha} y_{\alpha} + B_{ii} x_i^2
\end{eqnarray}
and the Gaussian integral can be evaluated to give
\begin{equation}
\Psi_i(x_i)\equiv\int\prod_{j\neq i}^N dx_j e^{-\frac{1}{2} \vec{x}^\intercal {\bf B} \vec{x}} \propto e^{-\frac{1}{2} \left[ B_{ii} - \sum_{\alpha=1}^{N-1} \frac{({\bf P}^\intercal \bf{b}_{\hat{i}})_{\alpha}^2}{\mu_{\alpha}} \right] x_i^2},\label{Psiixi}
\end{equation}
indicating that the distribution function of each node, measured individually, is also Gaussian whose variance depends on the network structure ${\bf Q}$ and the noise matrix $\boldsymbol{\sigma}$.

\section{Simulation results to illustrate ${\vec x}_{peak}\neq \vec X$ for a simple nonlinear NESS system}
\begin{figure}[H]
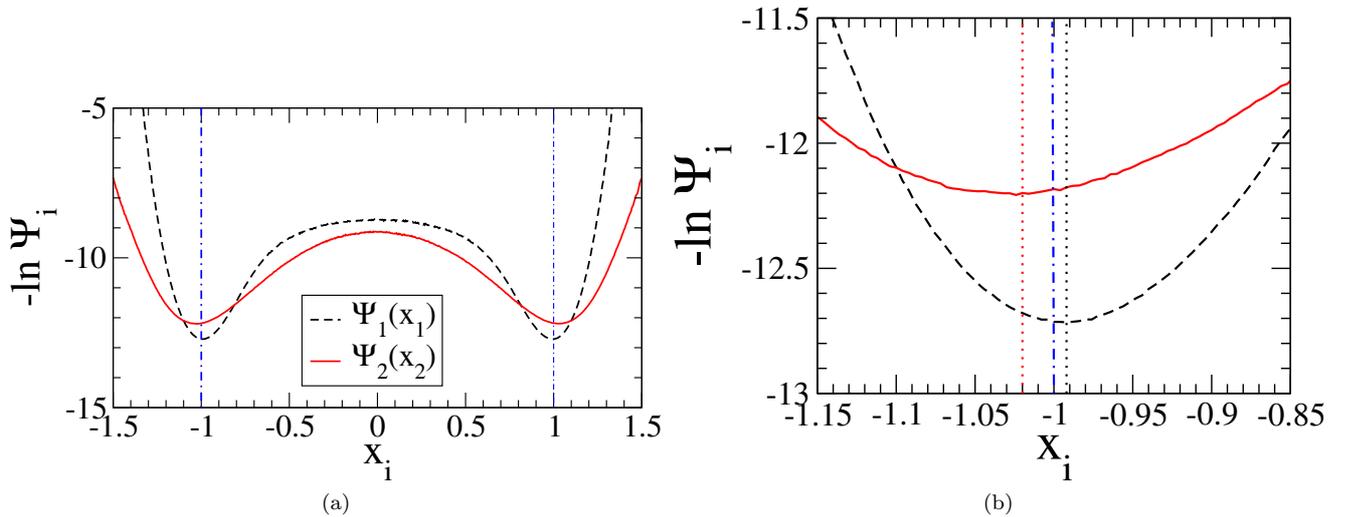

    \centering
    \subfigure[]{\includegraphics*[width=.48\columnwidth]{Fig12a.eps}}
    \subfigure[]{\includegraphics*[width=.48\columnwidth]{Fig12b.eps}}
    \caption{$N=2$ network with $f_i(x)=r_ix(1-x^2)$; $r_1=r_2=4$; $W_{12}=-W_{21}=5$; $\sigma_{11}=.2$, $\sigma_{22}=.25$  (a) $-\ln \Psi_1(x_1)$ and $-\ln \Psi_2(x_2)$. The noise-free stable fixed point values of $X_i=\pm 1$ are marked by vertical dot-dashed lines. (b) Magnification of (a) near $x=-1$ to show that the peak values of $\Psi_i(x_i)$ (the minimum value of $-\ln \Psi_i(x_i)$, which are marked by the vertical dotted lines,  are shifted from the noise-free value of -1.}\label{lnPsiN2}
    \end{figure}

Since $\vec \nu=0$ for a linear $\vec{\cal F}$, one expects $|\vec \nu|\sim {\cal O}(\delta x^2)$. The measured mean square fluctuations, $\frac{1}{N}\sum_{i=1}^N \langle \delta x_i^2\rangle\simeq 0.8\times 10^{-2}$ for the simulations in Fig. \ref{lnPsiN2}, which is of the same order as $|\vec \nu|\simeq 2 \times 10^{-2}$ as shown in Fig. \ref{lnPsiN2}b.
    
  \section*{Acknowledgements}
This work has been supported by the National Science and Technology Council of Taiwan under Grant  No. 113-2112-M008-018-MY2.

\bibliographystyle{unsrt}
\bibliography{references}

  \end{document}